\documentclass[onecolumn,usenatbib]{mn2e}

\usepackage{graphicx}
\usepackage{amssymb}
\usepackage{amsmath}
\usepackage{subfig}
\usepackage{comment}

\usepackage{color}
\usepackage{ulem}

\usepackage{color}
\usepackage{ulem}

\title[Collapse of protoplanetary clumps]{The collapse of protoplanetary clumps formed through disc instability: 3D simulations of the pre-dissociation phase}

\author[M. Galvagni et al.]{M. Galvagni,$^1$ T. Hayfield,$^2$
  A. Boley,$^3$ L. Mayer$^1$, R. Ro\v{s}kar$^1$ and P. Saha$^1$\\
  $^1$Institute of Theoretical Physics, University of Zurich,
  Winterthurerstr~190, 8057 Zurich, Switzerland \\
  $^2$Max-Planck-Institute fur Astronomie, Koenigstuhl 17, D-69117 Heidelberg, Germany \\
  $^3$Department of Astronomy, University of Florida, 211 Bryant Space Science Center, USA}

\begin{document}
\maketitle

\begin{abstract}
We present 3D smoothed particle hydrodynamics simulations of the collapse of clumps formed through gravitational instability in the outer part
of a protoplanetary disc. The initial conditions are taken directly from a global disc simulation, and a realistic equation of state is used to follow the clumps
as they contract over several orders of magnitude in density,
approaching the molecular hydrogen dissociation stage.
The effects of clump rotation, asymmetries, and radiative cooling are
studied. Rotation provides support against fast collapse, but non-axisymmetric modes develop and efficiently transport angular momentum
outward, forming a circumplanetary disc. This transport helps the clump reach the dynamical collapse phase, resulting from molecular hydrogen dissociation, on a thousand-year timescale, which is smaller than timescales predicted by some previous spherical 1D collapse models. Extrapolation to the threshold of the runaway hydrogen dissociation indicates that the collapse timescales can be shorter than inward migration timescales, suggesting that clumps could survive tidal disruption and deliver a proto-gas giant to distances of even a few AU from the central star.
\end{abstract}


\section{Introduction}

The combined efforts of RV and transit \citep[see][]{2012arXiv1205.2273W,2011ApJ...728..117B}, mircrolensing \citep{2011ESS.....2.0103S}, and direct imaging surveys \citep{2010ApJ...719..497L} have revealed a wide range of planetary systems, with planets on both very short and very wide orbits.  
These increasingly large sample-size distributions of known planets, although still biased, can be used to constrain planet formation theories. 
Unfortunately, testable predictions by planet formation theories are not fully developed, and in some cases, there may be multiple ways to produce a particular planet \citep{2010Icar..207..509B} or to induce the observed disc structure \citep[see][]{2012ApJ...748L..22M,2011ApJ...729L..17H}.
To complicate matters further, regardless of the planet formation mechanism, planet-planet and planet-disc interactions can lead to large scale transport of planets throughout their natal discs \citep[see][]{2011MNRAS.416.1971B,2011ApJ...737L..42M}.

There are currently two main theories \citep[for a review see][]{armitage} for the formation of massive planets: core accretion (hereafter CA) \citep{mizuno,pollack} and gravitational instability (hereafter GI) \citep{1997Sci...276.1836B,mayerquinn}. From these mechanisms, it is also possible to derive alternative scenarios, such as the tidal stripping theory \citep{2010Icar..207..509B,nayakshin}. In each paradigm, the formation of a gas giant planet proceeds in a very different way. In the first case the grains in the circumstellar disc aggregate until they reach a critical core mass at which there is a run away accretion of gas from the disc onto the solid core. The critical core mass for runaway gas accretion can vary due to conditions in the discs \citep{2010Icar..209..616M,2011ApJ...727...86R}; however, the large population of Neptune-mass planets on short periods discovered by {\it Kepler} suggests that the critical value may be $\gtrsim 10 M_{\oplus}$. Under the disc fragmentation paradigm, the giant planet is the result of a contraction of a massive self-bound clump of gas that forms due to fragmentation of GI-driven spiral arms.

The idea of forming giant planets by disc instability goes back to at least
\citet{1951PNAS...37....1K}. Renewed interest in disc instability as a planet-forming mechanism is largely  due to the advancement of  \citet{1997Sci...276.1836B} and to recent discoveries of extrasolar planets on wide orbits  \citep{bonavita,gemini,quanz}. The Toomre (1964) stability criterion measures the susceptibility of the disc to growing perturbations due to self-gravity; whenever
\begin{equation}
Q \equiv \frac{2c_s}{G\Sigma t_P}<1,
\end{equation}
axisymmetric ring distortions will grow in an infinitesimally thin disc. 
 Here $\Sigma$ is the disc surface density,
$c_s$ is the sound speed, and $t_P$ is the epicyclic period, which will be close to the orbital period in nearly keplerian discs. 
For three-dimensional protoplanetary discs with finite width, numerical simulations have shown that spiral arms will grow whenever $Q\lesssim1.7$ \citep{2007prpl.conf..607D}.
The mass scale for self-gravitating distortions (Toomre mass), which is not necessarily the fragmentation mass, can be expressed as a sound horizon
\begin{equation}
GM_T/c_s^2 \approx c_s t_P,
\end{equation}
which at plausible radii gives massive giant-planet to brown-dwarf masses within a square Toomre wavelength. Simulations have shown that when clumps do
form, initial fragment masses can be a factor of a few or more smaller than the Toomre mass \citep{2004ApJ...609.1045M,2010Icar..207..509B,rogers_wadsley_2012_mnras}. After spiral structure develops in a disc due to self-gravity, spiral arms could regulate the instability and reach a marginally unstable state, with mass redistribution and energy dissipation balancing the effects of disc cooling.  In the local limit,     
\citet{2001ApJ...553..174G} showed that for a cooling time $t_{\rm cool}$, GIs will lead to an effective \citet{shakura_sunyaev_1973} $\alpha$ viscosity given by
\begin{equation}
\alpha =   \frac 2{9\pi\Gamma(\Gamma-1)}  \frac{t_P}{t_{\rm cool}}
\label{crate}
\end{equation}
where $\Gamma$ is the (2D) adiabatic index. If GIs cannot stabilize the disc, then spiral arms can fragment into bound clumps. The cooling time boundary for fragmentation, which sets a maximum $\alpha$ in the local limit \citep{2005MNRAS.364L..56R}, is still being debated in the literature \citep{2011MNRAS.411L...1M,2011MNRAS.413.2735L,2012MNRAS.421.3286P}, but  appears to be within a factor of a few $t_{\rm cool}$. 

Now consider the conditions in a disc that are favorable to disc instability.  For $T=100$ K and $t_P\sim 12$ yr, $\Sigma\gtrsim 3600$ g cm$^{-2}$ for $Q\lesssim 1.7$. 
 For comparison, a minimum-mass solar nebula has a surface density $\Sigma \sim 150\, \rm g\, cm^{-2}$ at 5.2 AU, although we also remind the reader that that this value is a {\it minimum} and that the assumptions used to derive this mass are suspect \citep[for an example, see][]{2007ApJ...671..878D}.
For clump formation, higher surface densities than those considered above may be needed, as fragmentation seems to require an azimuthally averaged $Q<1.4$ \citep{2004ApJ...609.1045M}. The exact threshold will depend on the equation of state.  Because a low value of $Q$ is needed for fragmentation, it is of interest to know where low values will most likely persist.  In a Keplerian disc, $Q\propto r^{q/2-p-1.5}$ for disc radius $r$.  We have assumed that the temperature and surface density profiles can be described by power laws with indices $q$ and $p$, respectively.  For a flared disc with $q\sim -0.5$, the surface density must drop more steeply than $p\sim -1.75$ to prevent the outermost regions of the disc from being the most susceptible to GIs.  Furthermore, as the surface density drops and the disc becomes colder, the opacity of the disc will also drop and cooling can become more efficient, as long the gas has enough emissivity to radiate.  Because the local dynamical times are long at large radii, cooling times can be much shorter than the local dynamical time, and disc fragmentation, not just instability, becomes likely \citep[e.g.,][]{rafikov2009,boley2009}.

When clumps do form, they are many orders of magnitude less dense than a giant planet, with radii of a few AU, and have significant angular momentum. Although there have been detailed numerical studies of clump evolution, they have been limited to spherical quasi-equilibrium models \citep{2009ApJ...697.1256H}.
Simulations that follow the clump in 3D and within the context of a disc are too low resolution  to capture the internal dynamics of the clump \citep{2007A&A...475...37S}.  
As long as clumps remain large and diffuse, it has been suggested that they might lose a significant fraction of their mass and become super-Earths as long as they can accrete enough mass in the core \citep{nayakshin,2010ApJ...724..618B}. However, even prior to considering tidal mass loss, the actual mass that goes into forming the planet (or brown dwarf) and the mass that contributes to an eventual subdisc (circumplanetary or circum-brown dwarf) is unknown and depends on the efficiency of angular momentum transport within the clump. As discussed in more detail below, when the central temperatures of the clump reach $\sim 2000$ K, H$_2$ dissociation will cause rapid collapse of the clump, forming a bound substellar companion (gas giants and brown dwarfs). At this point tidal mass loss will be negligible due to a very high central density.

In the present study we present high resolution 3D simulations of clumps that form through disc fragmentation at $80$ AU. Their evolution is followed until the contaction approaches hydrogen dissociation and rapid collapse begins.  The simulations are performed using a new version of {\tt Gasoline} \citep{GasolineW} that implements an equation of state (herein EOS) designed to model the relevant temperature-density regime. In some of the simulations we also include the effect of radiative cooling. Section \ref{sec:met} describes our methods. Section \ref{sec:res} shows simulation results, and discussions and conclusions are presented in section \ref{sec:con}. A detailed derivation of the EOS is given in the appendix. The purpose of the present work is to follow the evolution of clumps in order to build a self-consistent picture of the types of objects that GI forms; this is a rich area of research, and this study represents an initial step toward this goal. 

\section{METHODS} \label{sec:met}

\subsection{EOS and Radiative Cooling}
\begin{figure}
\begin{center}
\includegraphics[scale = 0.4]{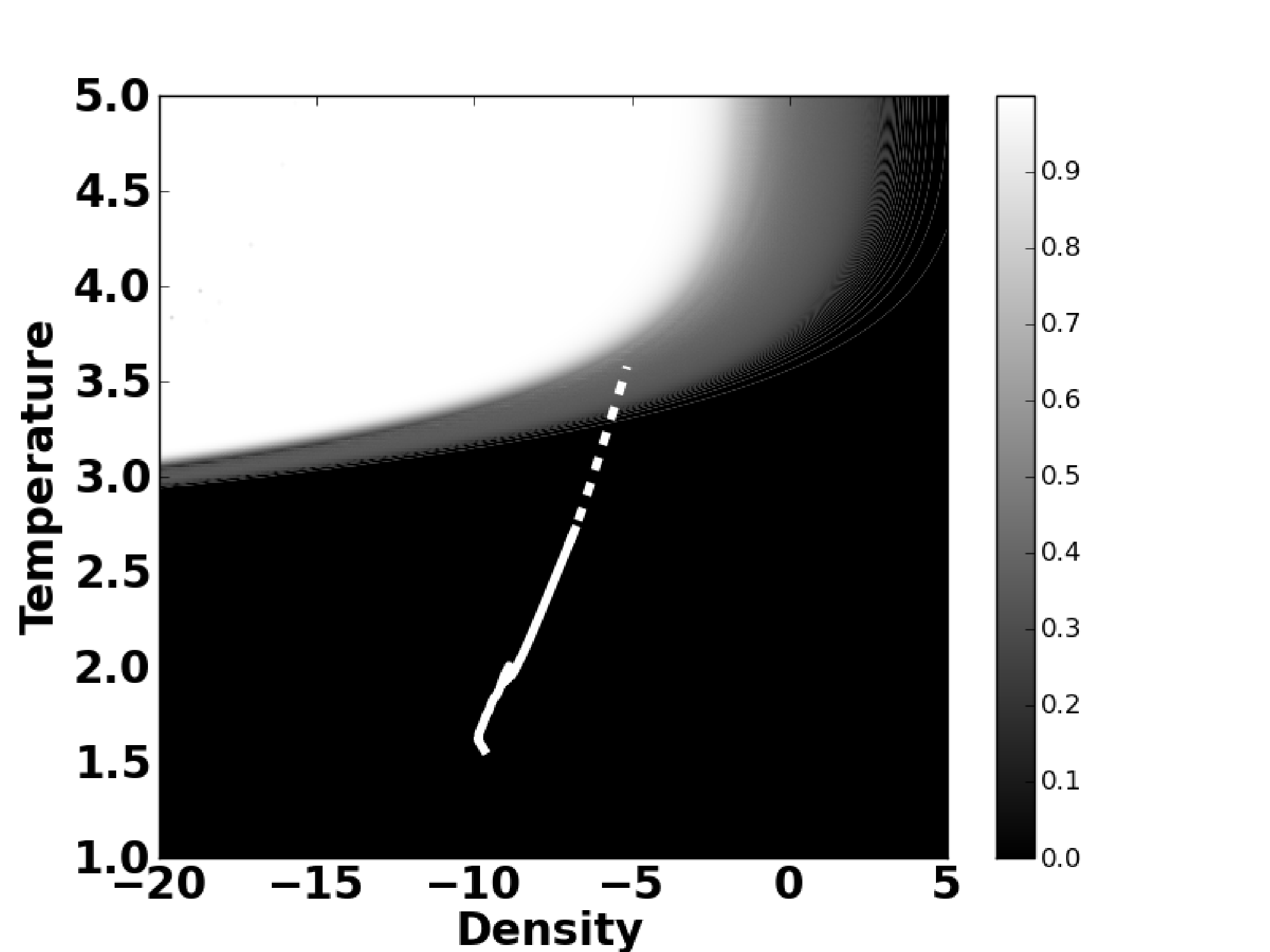}
\caption{Dissociation parameter {\tt a} as a function of temperature and
  density. The x and y axes are in log scale (cgs units). The white solid curve represents the evolution of the clump from the IC$1$ simulation in the cooling case, the white dotted curve is the extrapolated evolution.}
\label{fig:adiss}
\end{center}
\end{figure}

The internal structure of any single clump will go through a wide range of densities and pressures as the object cools and contracts, which can change the relative importance of the translational, rotational, and vibrational modes of molecular hydrogen during the clump's evolution. 
In particular, the onset of H$_2$  dissociation  in the clump's core will lead to dynamic collapse of the entire structure.
For these reasons, we have modified the SPH code {\tt Gasoline}
\citep{GasolineW} to include an EOS for a mixture of atomic
and molecular hydrogen in chemical equilibrium (see appendix A).
The EOS is taken to be
\begin{equation} \label{eos}
P = \frac{K_b T \rho (1+a)}{2 m_P},
\end{equation}
with $P$ being pressure, $k_B$ the Boltzmann constant, $T$ temperature, $\rho$ density, and $a = a(T, \rho)$ the dissociation parameter, given by the number of protons in the atomic case over the total number of protons, and $m_P$ proton mass. In Figure (\ref{fig:adiss}), we show the dissociation parameter as a function of density and temperature.   Dissociation becomes more likely as the temperature increases and as the density decreases. In order to highlight the regime of interest in this context, the evolution of one of our simulations is also shown.

For a clump to evolve from the initial low-density, low-temperature state to the H$_2$ dissociation threshold, it must radiate away energy. 
In the following, we present simulations with and without cooling. The latter is referred to as the adiabatic case, and serves as a base for comparisons.  We implement cooling in {\tt Gasoline} for these simulations with a simplified prescription that uses local gas conditions only \citep[as done in][]{2010Icar..207..509B}. Let the energy loss per time per volume be given by  
\begin{equation} \label{cool}
\Lambda = (36 \pi)^{1/3} \frac{\sigma}{s} (T^4 - T^4_{\mbox{min}}) \frac{\tau}{\tau^2 + 1}
\end{equation}
where $\tau$ is the optical depth, $s=(m/ \rho)^{1/3}$ and $\sigma$ is the Stefan-Boltzmann constant.  Here, a minimum background temperature $T_{\mbox{min}} = 10$ K is assumed, which is the reference background radiation field for the ISM. The clump explored here forms in the outermost regions of the disc, where the tempertature can become dominated by the radiation background, which is what we assume. The optical depth is given by $\tau=\rho \kappa s$, where the opacity is approximated as 
\begin{equation}\label{opacity}
\kappa = \sqrt{\frac{T'}{64}}~\rm cm^{2} g^{-1}
\end{equation}
for $T' = T$ if $T > T_{\mbox{min}}$ and $T' = T_{\mbox{min}}$ otherwise.  While equation (\ref{cool}) is only approximate, it allows us to capture the general behavior of radiative cooling. Namely, cooling is most efficient at an optical depth $\tau \sim 1$.  The opacity law used here is also very approximate, but is chosen to be monotonically increasing with temperature for two reasons: (1) Only a local optical depth is used, where the proper optical depth is integrated along a path. (2) The geometry of the problem places the densest material at the highest temperatures.  Although sublimation of ices, organics, and dust can all cause the opacity to drop suddenly, the integrated optical depth for radiation to leave the clump from any of these regions will likely be large.  The simple form for the opacity permits fast evaluation of radiative cooling, allowing us to focus on investigating the effects of the EOS on the clump's evolution.  The effects of radiative transfer in addition to the new EOS will be investigated in future work.

\subsection{Initial conditions} \label{sec:ics}

\begin{table}
\caption{Simulations parameters for the global simulation presented in Boley et al. 2010 and for the high resolution simulations presented in this work.}
\begin{center}
\begin{tabular}{c|c|c|c|c|c|c|}
\hline
 & Mass [$M_J$] & Radius [AU] & N. particles & Grav.Softening & N. neighbours\\
 \hline
 Global Simulation & 3.7 & 7.0 & 1.5 $\times 10^4$ & 0.5 AU & 32 \\ 
 Clump Simulations & 3.7 & 7.0 & 1.5 $\times 10^5 $ & 0.045 AU & 32\\
\hline
\end{tabular}
\end{center}
\label{tab:sim}
\end{table}

We present four high-resolution simulations of gaseous clumps that have formed through disc fragmentation. Clump initial conditions (ICs) are taken directly from global simulations of a fragmented protoplanetary disc \citep{2010Icar..207..509B}, but with the resolution increased by a factor of $10$ (see below for a description of the procedure). The clump was extracted from the global simulation when the fragment's central density was one order of magnitude larger than the background value. Its radius was taken to be twice the bound radius. We selected a clump that has a stellar separation of about $80$ AU and is not distorted due to tidal effects from other close clumps. This choice ensures that the clump has a simple morphology and that it is not strongly affected by the presence of other objects or the central potential (see Figure \ref{fig:roche}). By using ICs based on the results of global simulations, we are confident that the initial conditions of the presented simulations are self-consistent with the formation of clumps by disc instability. We refer to the ICs that are produced by this direct extraction method as IC$1$. As the aim is to follow the collapse of the object through several orders of magnitude in density, the simulations are quite computationally expensive.

We increased the effective resolution in the extracted clump by resampling the ``parent'' particles (i.e. those taken from the global simulation) by ten ``child'' particles, randomly distributed within a volume defined by the SPH kernel smoothing length $h$ determined using 16 neighbors. The child particles inherited equal fractions of the mass from their parents as well as their velocity. However, in order to preserve the multiphase structure of the clumps, the hydrodynamic quantities were distributed to the child particles using a standard SPH scatter scheme.The temperature has been rescaled to take into account the change in the mean moleclar weight $\mu$ and in the adiabatic index $\gamma$ between the global simulation and the high resolution ones used in this study. To simplify the software development, we reused the neighbour-finding methods of the well-known group-finding code { \tt SKID} \citep{Stadel:2001} and tailored them to our needs. 

The ICs produced by direct extraction can be used to produce a second, quieter set of ICs. This second set (IC$2$) is produced by symmetrizing the mass and the temperature profiles of IC$1$ and by setting the initial velocity field, which is mostly rotational, to zero. This gives us a spherically symmetric IC that can be used to isolate the effects of rotation on clump contraction. For each IC, one simulation is run adiabatically (no cooling) and a second is run with the radiative cooling described in section 2. The adiabatic case will help to isolate the EOS effects from clump cooling during the early stages of the system evolution. 

We check whether the tidal potential of the central star can affect the clump evolution by running an adiabatic simulation with IC$1$ for $\sim 10$ of its internal dynamical times with the central star explicitely included. We found negligible differences between clump evolution with and without the central star. At the time that the clump was extracted from the global simulation, the clump is at large stellar separation and well within its Hill sphere. Tidal effects only play a role in the outer, low-density layers of the clump, which appear to be unimportant for the evolution of the clump as studied here (see section \ref{sec:de} for a detailed analysis).

The extraction method used to create the initial conditions prevents us from studying the effects of gas accretion from the disc onto the clump during its evolution. This is a limitation for these simulations. Gas accretion potentially plays an important role in the evolution, although it is not clear in which direction it would lead. As a first approximation, it would speed up the collapsing timescale as it increments the clump mass. At the same time, though, this process would influence the density and the temperature of the object, leading to a different cooling rate which is not easy to predict due to the non-linear nature of cooling. In order to correctly quantify the effects of gas accretion, then, long term simulations on large scale are needed, which are beyond the aim of this work.

Table \ref{tab:sim} summarizes the numerical parameters used in this study, including the initial clump mass and radius. Figure \ref{fig:ICprof} shows the temperature, density and cumulative angular momentum profiles of the initial clump resampled at higher resolution (see next section). The angular momentum barrier $R_b$ is the radius that the object would have if it were rotationally supported, where $R_b(<r) = J(<r)^2 / GM(<r)$ with specific angular momentum inside r $J(<r)$. As the initial radius is $2.5$ AU and the corresponding angular momentum barrier is $0.17$ AU, the clump is only partially rotationally supported. This is confirmed by the initial ratio between rotational and gravitational energy, which is $0.18$.

Figure \ref{fig:vir} shows the different components of the energy as a function of the radius for the initial condition (kinetic, thermal and gravitational). It is evident that the the gravitational component is dominant and that the clump is out of equilibrium. The virial equilibrium condition $2(E_{\mbox{kin}} + E_{\mbox{the}}) = -E_{\mbox{gra}}$ \footnote{Note that only the translational component of the thermal energy goes into pressure support. If the total thermal energy is used, then the virial condition is dependent on the volume-averaged adiabatic index, such that $2E_{\mbox{kinetic}} + 3 (\bar{\gamma}+1)E_{\mbox{thermal}} = -E_{\mbox{grav}}$.} is not completely fulfilled, as the initial condition is missing $4 \%$ of the internal energy needed for it to be in virial equilibrium. This out-of-equilibrium condition is due to the low resolution of the global disc simulation from which we are extrapolating the clump, as in that condition the clump was not allowed to properly collapse because of a gravitational softening of $0.5$ AU. This artificial out-of-equilibrium state leads to a fast initial collapse phase, which lasts for $4 t_{\mbox{dyn}}$. The clump evolution is then self-consistent after this initial transient.

\subsection{Determining a Resonable Clump Resolution} \label{sec:res}
The resolution of the clump as taken from the global simulation has been increased by a factor of $10$ through particle splitting. This value of $10$ has been selected based on the results of a resolution study, where IC$2$ has been evolved adiabatically at different resolutions. The initial temperature is kept constant for all particles that are split from their parent. The new particle mass is rescaled so that the total mass is constant between different resolution runs and the softening ($h_{\mbox{hsm}}$) is rescaled so that $h_{\mbox{hsm}} \times (N_{\mbox{par}})^{1/3}$ is constant. Figure \ref{fig:reso} shows the evolution of the half-mass radius and of the inner density in different runs; there is convergence for both quantities at increasing number of particles. The temperature profiles for three runs at different resolutions at different times are shown in Figure \ref{fig:treso}. In this case, the convergence at high resolution is clear. Moreover, in the low-resolution case, the inner part of the clump does not reach as high of temperatures as found in the high-resolution runs, as the simulation is not  resolved well enough to follow the collapse.

\section{RESULTS}

Using the simulations presented here, we follow the evolution and fate of clumps as they evolve toward second core formation. This evolution is strongly correlated to the rotational state of the object, so our analyses first focus on the effects of rotation and angular momentum transfer due to the growth of structures in the clump. These processes affect the evolution of the clump, which we quantify by studying its half-mass radius, inner density and temperature for the simulated timescale. Lastly, we give an estimate of the clump entropy and compare it with the values usually assumed in long-term contraction and cooling simulations.

\subsection*{Radius and angular momentum evolution}
We follow the evolution of the clumps in four different cases: IC$1$ and IC$2$ are evolved using the adiabatic and cooling cases. The evolution of the half-mass radius (Figure \ref{fig:hmr}) shows that the initial rotation of the clump (IC$2$) helps to prevent the collapse. This is a trivial result, but note how this effect is quite drastic and fast. In only a few dynamical times the difference between the two initial conditions is evident. Naturally, including a cooling term exacerbates the initial collapse.

The presence of initial asymmetries in the IC$1$ simulations leads to the development of Fourier modes, which are due to low-amplitude spiral structure. They are presumably initially seeded by the self-gravity of the disc and then amplified during the collapse phase by the clump dynamics. The stellar potential does not play a role during this phase of the evolution.

The strengths of the asymmetries are given by the global Fourier amplitudes, 
\begin{equation}
A_m = \frac{(a_m^2 + b_m^2)^{1/2}}{\pi\int \rho_0 \bar{ \omega} d \bar{\omega} dz} ,
\end{equation} 
with
\begin{eqnarray}
a_m &=& \int \rho \cos ( m \phi) \bar{ \omega} d \bar{ \omega} dzd \phi ,~\mbox{and} \\
b_m &=& \int \rho \sin ( m \phi) \bar{ \omega} d \bar{ \omega} dzd \phi .
\end{eqnarray}
Integration is extended out to $7$ AU. From this analysis it is clear that the clump develops an m-2 mode, which becomes weaker with time (see Figures \ref{fig:m2a}, \ref{fig:m2c} and \ref{fig:mtime}). These structures move the angular momentum outward, as can be seen in Figure \ref{fig:sp}, where, after the fast initial collapse, the cumulative angular momentum profile appears to be flattened in the inner part. When the modes start to become unimportant, this angular momentum transport becomes weaker, with the clump developing a spherical core (with only $46 \%$ of the initial mass collapsed into the core in the cooling case) surrounded by a rotating circumplanetary disc. In the adiabatic simulation, although the m-2 mode lasts for a longer time, the effects due to the angular momentum transfer are weaker compared to the cooling case. The total angular momentum is conserved in both simulations up to $5 \% $\footnote{The conservation of the total angular momentum can not be seen from Figure \ref{fig:sp} as the x-axes are only up to $1.5$ AU, so the component from the outer part is not shown.}.

Figure \ref{fig:tw} shows the evolution of the ratio between rotational and gravitational energy T/$\|$W$\|$ for IC$1$ in the adiabatic and cooling cases. In both cases this ratio initially increases, as the collapse is very fast and wins over the angular momentum transport. After the first few dynamical times, the collapse becomes slower and the m-2 modes move angular momentum outward. Eventually the T/$\|$W$\|$ stops increasing and instead decreases. This phase is faster in the adiabatic case as the clump's collapse is less efficient. It is worth pointing out that the initial condition IC$1$ is not spherical, so the values of  T/$\|$W$\|$ are not a straighforward indicator of the modes evolution of the clump. Indeed, we detect the presence of several fourier modes throughout the collapse, although the ratio hardly reaches $0.274$, which is the threashold value for a bar mode in a spheroidal \citep{1986ApJ...305..281D}.

While the clump is largely a spheroid, it can also be decomposed into core-like and disc-like structures. For the disc portion, the Toomre Q ($1964$) parameter can be used to explore the susceptibility of the system to the growth of non-axisymmetric structure. However, this must be taken with caution, as portions of the disc have significant pressure gradients, making a strict application of the Toomre Q difficult. Figure \ref{fig:toom} shows the evolution of Q in the cooling case IC$1$ outside the half-mass radius for five different times. Although early in the clump's evolution the clump's Q value approaches the instability threshold for a thick, $3$D disc (e.g. \cite{2004ApJ...609.1045M}), the disc evolves toward a stable state.

At the final stage of the simulation, the protoplanetary disc can be defined as the region between the clump radius and one third of the Hill radius, which is the maximal disc extent found in previous works \citep{2009MNRAS.397..657A,2011MNRAS.413.1447M}. Although these previous works address clumps formed via CA, with gas accreting onto solid cores, the disc which forms in the simulation herein presented has very similar properties. Figure \ref{fig:disc} shows the disc's morphology; the disc gas mass is $46 \%$ of the total mass, giving the clump and disc having comprable mass. Moreover, the clump velocity is sub keplerian in the core-like structure due to pressure support, while it approaches keplerian values in the disc-like structure. The ratio $H/r$ where $H$ is the disc hight is comparable to the values found in previous works.

\subsection*{Density and temperature evolution} \label{sec:de}
To understand if a clump is going to survive inside a disc, it is essential to study the evolution of the clump's inner temperature and density, where we use inner to refer to the values that are averaged over the particles inside one gravitational softening from the center of the clump. This average is used because the gravitational softening is larger than the particles' smoothing length. This also implies that the presented simulations underestimate the contraction, regardless of their high-resolution.
The evolution of inner density and temperature are shown in Figures \ref {fig:dens}, \ref{fig:temp}, \ref{fig:densE} and \ref{fig:tempE}. The density profile evolution in both the adiabatic and cooling cases shows that a fraction of the mass lies outside the Hill radius ($13.1$ AU in the initial condition), which will be stripped away due to the interaction with the host star, reducing the mass of the clump. This process will happen on a timescale comparable to the rotational time of the material at the Hill radius, which is (for our IC$1$) about $760$ yr. This timescale is longer than the duration of the simulations, so neglecting the star is not expected to alter the evolution of the high-resolution simulations. However, the timescale is shorter than the contraction timescale, which means tidal effects could still play some role in the evolution of the clump over the contraction timescale.  This effect is expected to be small for the conditions considered here, as the clump becomes highly concentrated over the duration of the simulation.

The simulations with cooling show that the clump will eventually reach the second core collapse. It is not possible to state  with certainty the same for the adiabatic simulations, as it is not clear which physical phenomena will lead to the collapse after the removal of the m-2 mode. Without cooling, the clump should eventually reach an equilibrium. In the cooling case, it is possible to estimate the timescale for reaching the dissociation of molecular hydrogen\footnote{In this work we consider the clump to reach the second core collapse when the dissociation parameter is $10 \%$ in the inner part of the clump. This is a safe assumption as the run away process due to dissociation of molecular  hydrogen actually starts when { \tt a} is only a few \%.} by extrapolating temperature and density (the first quantity is extrapolated linearly, while the latter using a parabolic function). For IC$1$ the extrapolation is done using the last $15$ $ t_{\mbox{dyn}}$, while for IC$2$ it is done using the last $1.5$ $t_{\mbox{dyn}}$. 
Although it is initially unclear whether such extrapolation is reasonable, as after the 3D simulations end, the clumps must still go through a wide range of temperatures and densities before collapsing to the second core, the simple extrapolation does place the inner values on a sensible trajectory (see Fig.~1). The speed of collapse will be dependent, in part, on the adopted cooling approximation for radiative cooling, but also on the opacity and metallicity of the clump, where for a given metallicity the opacity is only known to factors of a few.  These extrapolations are crude, but we expect the resulting timescales to be within the range of plausible evolution scenarios for realistic clumps.  See table \ref{tab:estimate} for the estimated timescales and physical quantities.

\begin{table}
\caption{The last density of the clump as taken directly from the simulations, as well as extrapolations for two evolutionary timescales and two thermodynamic values just priori to dissociation and collapse. See the text for further details.}
\begin{center}
\begin{tabular}{|c|c|c|c|c|}
\hline
Quantity & IC$1$: adiabatic & IC$1$: cooling & IC$2$: adiabatic & IC$2$: cooling \\ 
\hline
Last density [g/cm$^3$] & $4.20 \times 10^{-9}$ & $1.57 \times 10^{-7}$ & $4.28 \times 10^{-7}$ & $4.30 \times 10^{-6}$  \\
$1300$ K Timescale [yrs]  & $ 2.3 \times 10^{4}$ & $ 2.2 \times 10^{3}$ &  -- & $1.38 \times 10^{2}$\\
Dissociation Timescale [yrs]  & -- & $6.3 \times 10^{3}$ &  -- & $6.3 \times 10^{2}$\\
Extr. density [g/cm$^3$] & -- & $7.3 \times 10^{-6}$  & -- & $1.3 \times 10^{-4}$\\
Extr. Specific Entropy [$k_B$] & -- & $14.8$  & -- & $15.2$\\
\hline
\end{tabular}
\end{center}
\label{tab:estimate}
\end{table}

\subsection*{Entropy evolution}
It is usually assumed that CA and GI lead to the formation of protoplanets with different properties. In the first case, called the cold model ($s/k_B < 10$), these objects are supposed to have a lower specific entropy than in the second case, the so called hot model ($s/k_B > 10$) \citep{burrows}. With our simulations we are able to give an estimate of the actual value of the specific entropy when the clump reaches $R = 10 R_J$ by extrapolating  the entropy evolution (see table \ref{tab:estimate}). The specific entropy of the clump changes very slowly with time, and the extrapolated value agrees with the hot model range found in previous works. 

The evolution of specific entropy changes depending whether the inner or the outer part of the clump is considered. It is interesting to notice that, in all the performed simulations,  the specific entropy of the inner part of the clump decreases with time, while it increases for the outer part. This points towards the existence of a redistribution mechanism of the entropy between the clump and its envelope. Figure \ref{fig:entrE} shows the evolution of the specific entropy profile for IC$1$ in both the adiabatic and cooling case. Overall redistibution is evident, although the simulation with cooling develops a more complex specific entropy profile with a peak and a trough at $r_p = 0.48$ AU and $r_t = 2.38$ AU, respectively. This effect is due to the more effective collapse seen in the simulations with cooling, so that at $r = r_p$ the density has significantly decreased compared with the adiabatic case (factor of $5$), which leads to a local entropy increase.

\section{Discussion and Conclusions} \label{sec:con}

Gravitational instabilities in the outer regions of circumstellar discs can lead to disc fragmentation and the formation of clumps, but at the moment it is unclear whether these clumps can survive to become gas giant planets or other bound objects.  In this work, we have analyzed the initial collapse of realistic clumps using a set of high-resolution, 3D simulations that include an EOS that is appropriate for clump densities and temperatures.  Our results allow us to estimate the long-term evolution of these clumps, which can be used to address their survival and whether they can continue to contract to form a companion (in this case, a gas giant planet).

Our timescale estimates for the collapse are usually smaller than the values found in previous work. In \citet{helled} the timescale estimated to reach an inner temperature of $1300$ K (value at which the dust component evaporates) is $1.6 \times 10^4$ years for a clump with $3 M_J$. It is possible to make a comparison between this case and our IC$1$ simulation in the cooling case as it implements a similar physics and as the initial conditions have comparable values for both mass and luminosity\footnote{The luminosities are evaluated at half-mass radius in order to exclude the circumplanetary envelope. The initial value is $4.5 \times 10^{29} \mbox{erg/s}$, while the end value for the IC$1$ simulation with cooling is $4.0 \times 10^{28} \mbox{erg/s}$.} \citep[see][]{2011Icar..211..939H}. Our result for estimated timescale to reach the same inner temperature shows that a 3D treatment of the clump evolution describes a faster collapse (see Table \ref{tab:estimate}). This difference is due to a combination of factors, above all the different treatment for contraction. Indeed \citet{helled} implements a 1D quasi-static model, which means that the collapse goes through a series of equilibrium states. In this way if reactions take place on a dynamical timescale that is smaller than the sound crossing time, the shells are not able to communicate and each of them has to wait for the evolution of the others to react to it, while non-axisymmetric and dynamical instabilities can occur in our 3D simulations. 

The estimated timescales presented in this work have important implications for the evolution and survival of clumps formed by disc instability. The clump contraction timescales and central mass concentrations determine whether fragments can survive tidal stripping forces as they move throughout the disc. It is possible to estimate the minimum distance from the central star that the clump can reach before tidal forces will overwhelm the clump's stability and prevent collapse to a second core.  To calculate this, use the Hill radius definition:
\begin{equation}
a_{\mbox{min}} = R \left( \frac{3M_{\mbox{star}}}{m_{\mbox{clump}}} \right)^{1/3}
\end{equation}
with $M_{\mbox{star}} = 0.3 M_{\odot}$. Using the results from the more realistic simulation (IC$1$ with cooling), i.e., the values from the last output, we find $a_{\mbox{min}} = 5.97$ AU. It is also possible to extrapolate the half-mass radius of the clump to its value as the clump reaches collapse to a second core by $R = R_{\mbox{last}} \left( \rho_{\mbox{last}} / \rho_{\mbox{extr}} \right) ^{1/3} = 0.32$ AU, so that $a_{\mbox{min}} = 2.52$ AU. This means that if the clump collapses before getting closer to the star than $a_{\mbox{min}}$, it can become bound enough to survive further tidal effects. This is expected to be the case, as the migration timescale for such an object has been found to be of the order of $10^4$ yr  \citep[compare][]{2011MNRAS.416.1971B,2011ApJ...737L..42M}, which is  longer than the extrapolated time to molecular hydrogen dissociation in the clump. There is the possibility that the clump gets disrupted at pericenter passages in the early stage of migration through clump-clump scattering, clump-spiral arm excitation, or birth on an eccentric orbit. If migration remains smooth, then the clump can remain well inside its Hill sphere for plausible migration rates.

Although rapid collapse could be a common evolutionary scenario for clumps, there is still the potential for rich dynamics to occur prior to molecular hydrogen dissociation. This is best illustrated by noting the total angular momentum of the IC1 clump with cooling at the end of the simulation, which is $L = 5.4 \times 10^{47}$ g cm$^2$ s$^{-1}$. This result is in agreement with the calculations shown in \citet{machidaAM} which give a slightly larger value for $L$ (less than factor of $2$) for a clump with the same mass.  In both cases, the total $L$ is two orders of magnitude larger than the angular momentum estimated for Jupiter ($L_{j} = 4.14 \times 10^{45}$ g cm$^2$ s$^{-1}$). Moreover, as the clump evolves, the amplitude of the fourier modes get weaker in these simulations, leading to a less efficient removal of the angular momentum from the clump. This implies that there has to be a second mechanism, later in the evolution of the clump, in order to match the simulation results for $L$ with Jupiter. This mismatch could, for example, be due to the neglect of magnetic effects.  The temperature and density regimes that the clump will experience do allow for possible development of some magnetic drag, due to thermal ionization, that would lead to angular momentum loss \citep[compare with][]{rosalba}. The presence of magnetic fields in the protoplanet could also lead to a transfer of its angular momentum to a circumplanetary disc due to coupling of the planetary dipole field lines to the disc fluid, as discussed in \citet{takata}. Although the Takata and Stevenson derivation applies only for the very last stage of the planet formation, when its radius is only a factor of ten larger than its final value, the mechanism they describe can lead to the decreasing of $L$ by a factor of $3-4$ in this very last phase. Self-gravitating instabilities could also become important again as the clump contracts and the core spins up.  If $T/\vert\vert W\vert\vert$ once again reaches $\sim 0.27$, then dynamic bar instabilities can be rejuvenated. 

The simulations that have been performed allow us to confirm that the clumps formed via GI are described by the so called hot state model. As stressed in \citet{burrows}, this is particularly interesting as it can allow the next generation of observational surveys to discriminate between the GI and the core accretion model.

The present work points toward a more central role for the GI theory in the direct formation of gas giant planets, as the contraction of disc instability clumps may be very rapid, at least for some if not many conditions. The possibility that protoplanetary clumps retain most of their mass even when they reach distances close to the star as a result of inward migration opens new scenarios for the origin of close-in extrasolar plantes. Indeed, while previous work has suggested that partial stripping and concurrent core formation could turn clumps into super-Earths and/or Neptunes \citep{2010Icar..207..509B,nayakshin}, our findings suggest that in principle a fraction of clumps formed at large radii by GI can become giant planets on close-in orbits. Nevertheless, it is worth pointing out that this result comes from extrapolated values, and not from a self-consistent study of a clump's contraction over the full range of its first core phase. This work is a first step towards a full description of clump contraction, which appears to be a very complex and computational demanding  problem. Note also that the EOS herein implemented does not really affect the clump evolution in the presented simulations, but it is expected to play a major role in the later evolution. A more realistic cooling prescription using radiative transfer, which is only mimicked here, as well as simulations taking into account gas accretion from the disc onto the clump, will need to be included in future models. These points will be addressed in future work.

\section{Acknowledgments}
The authors thank J. Wadsley, the main author of the { \tt Gasoline} code, for useful advices and F. Meru for helpful discussions.

\clearpage 

\begin{figure}
\centering
\subfloat{\includegraphics[width = 0.5\textwidth]{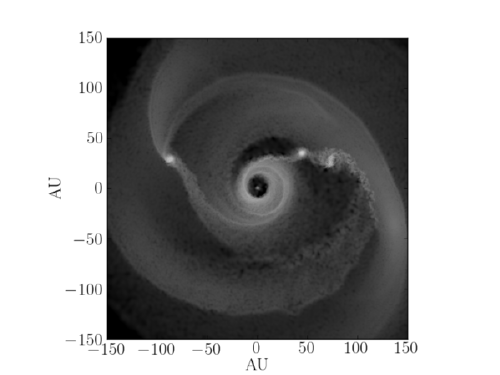}}
\subfloat{\includegraphics[width = 0.5\textwidth]{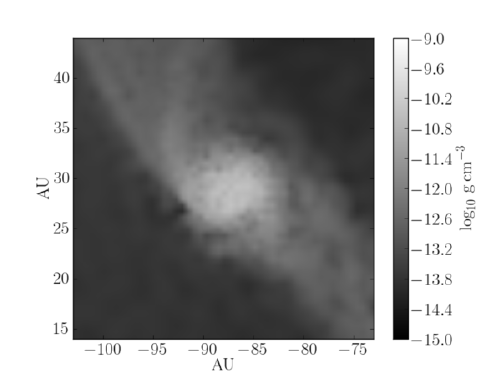}}
\caption{Density map for the initial condition. On the left, the propoplanetary disc simulation presented in Boley et al. 2010. On the right, a zoom on the selected clump. Axes are in AU, density in cgs (log scale).}
\label{fig:ic}
\end{figure}

\clearpage 

\begin{figure}
\centering
\subfloat{\includegraphics[width = 0.5\textwidth]{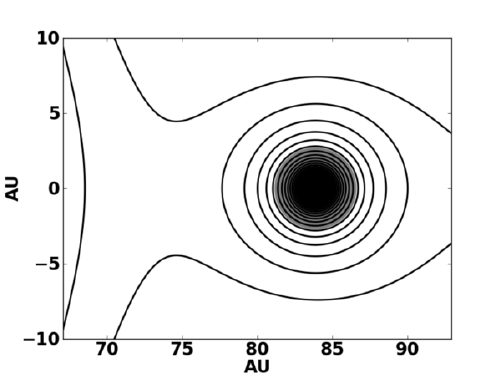}}
\subfloat{\includegraphics[width = 0.5\textwidth]{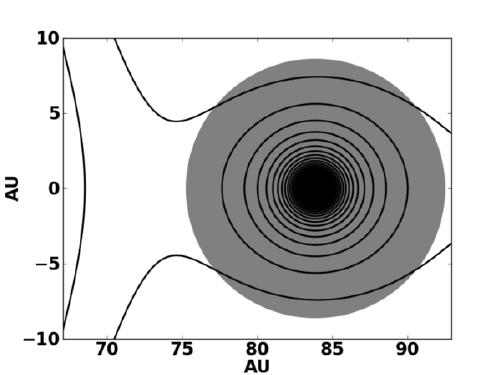}}
\caption{Roche map of the protoplanetary disc. The black curves represent the gravitational potential, the gray circle
  the extracted clump at half mass (left) and $80 \%$ mass
  (right). Axes are in AU.}
\label{fig:roche}
\end{figure}

\begin{figure}
\centering
\subfloat{\includegraphics[width = 0.3\textwidth]{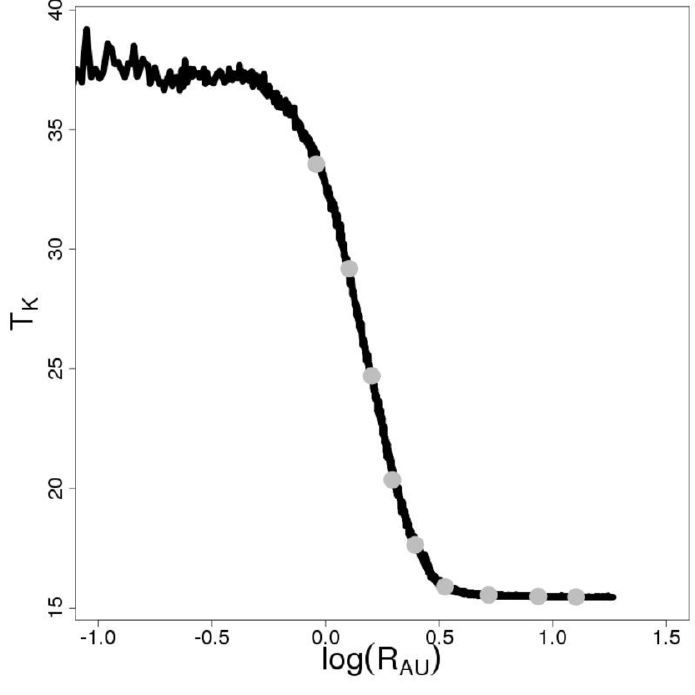}}
\subfloat{\includegraphics[width = 0.3\textwidth]{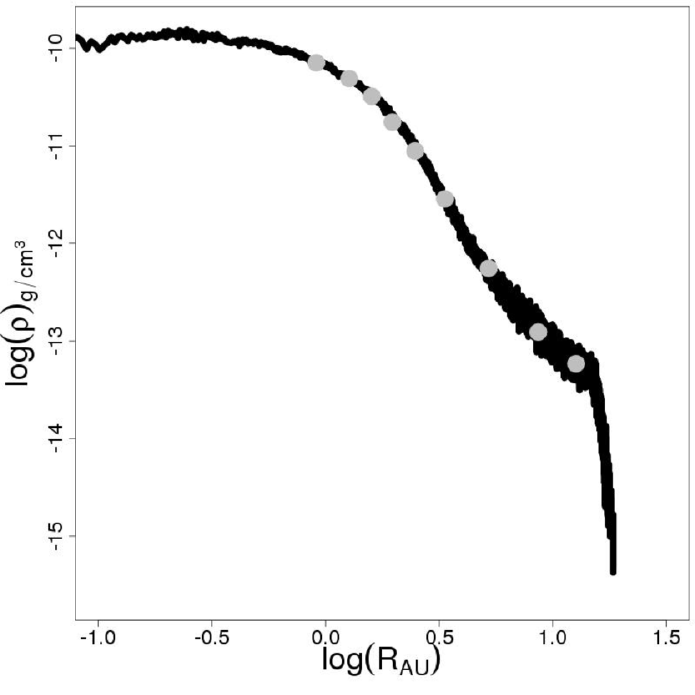}}
\subfloat{\includegraphics[width = 0.33\textwidth]{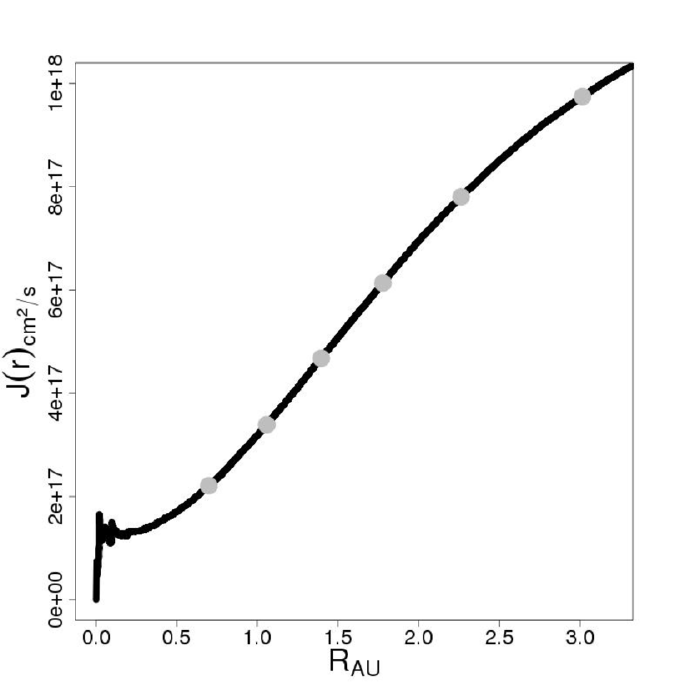}}
\caption{Temperature (on the left), density (on the middle) and cumulative angular momentum (on the right) profile
  of the initial clump. The gray dots represents shells containing $10 \%$ of the mass.}
\label{fig:ICprof}
\end{figure}

\clearpage

\begin{figure}
\centering
\includegraphics[width = 0.4\textwidth]{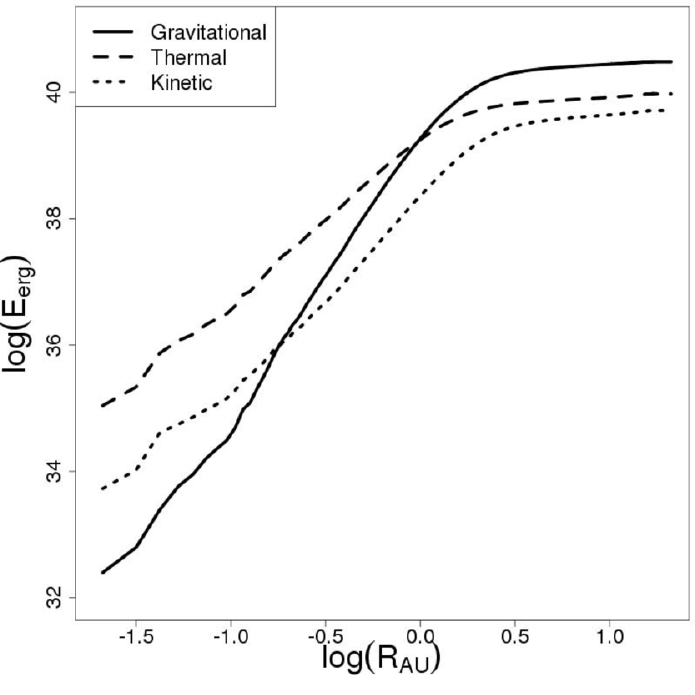}
\caption{Cumulative energies (gravitational, kinetic and thermal) in log scale (cgs units) as a function of the radius (in AU) for the initial condition.}
\label{fig:vir}
\end{figure}

\begin{figure}
\centering
\subfloat{\includegraphics[width = 0.5\textwidth]{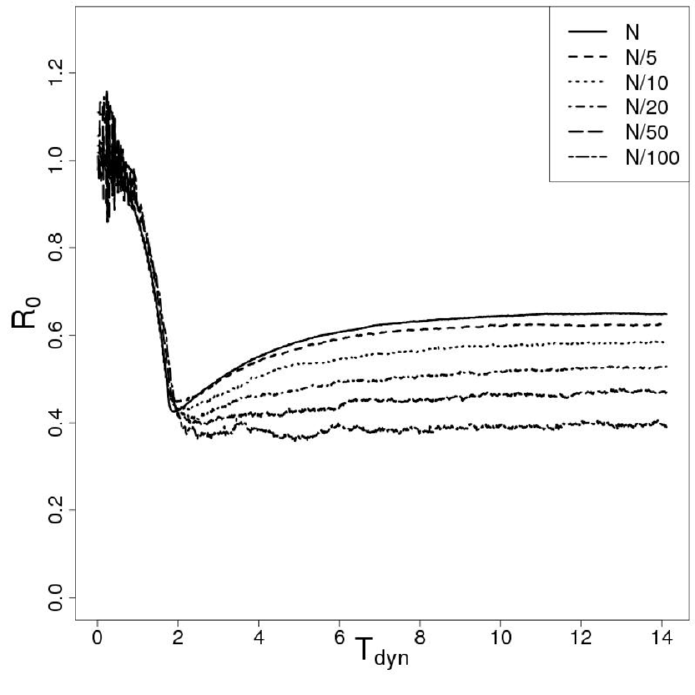}}
\subfloat{\includegraphics[width = 0.5\textwidth]{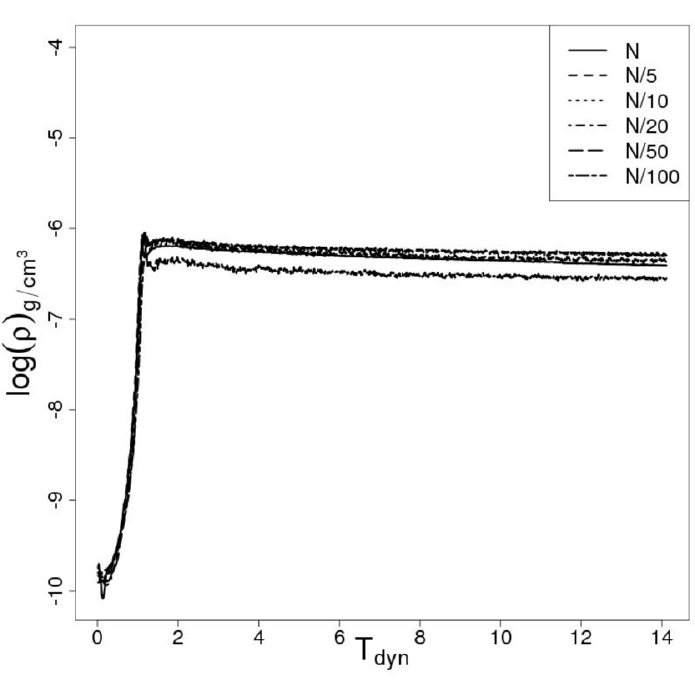}}
\caption{Evolution (in dynamical times) of the half-mass radius (in
  unit of its initial value) and inner density (in
  log of cgs) for IC$2$
  in adiabatic case for different number of particles. N is the
  number of particles in the high-resolution case.}
\label{fig:reso}
\end{figure}

\clearpage 

\begin{figure}
\centering
\subfloat{\includegraphics[width = 0.3\textwidth]{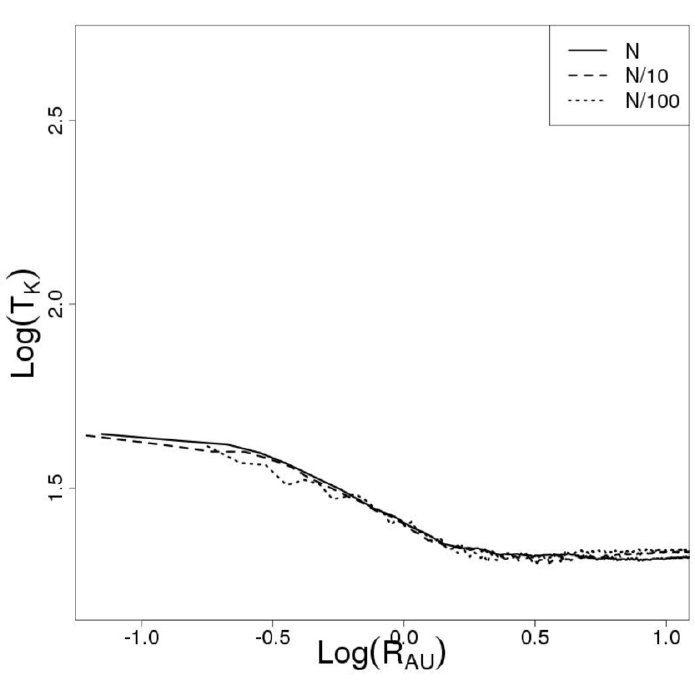}}
\subfloat{\includegraphics[width = 0.3\textwidth]{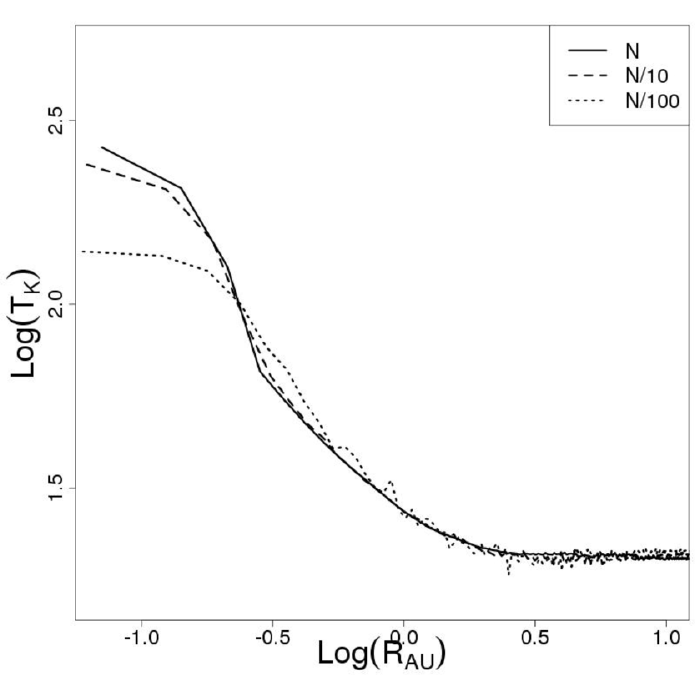}}
\subfloat{\includegraphics[width = 0.3\textwidth]{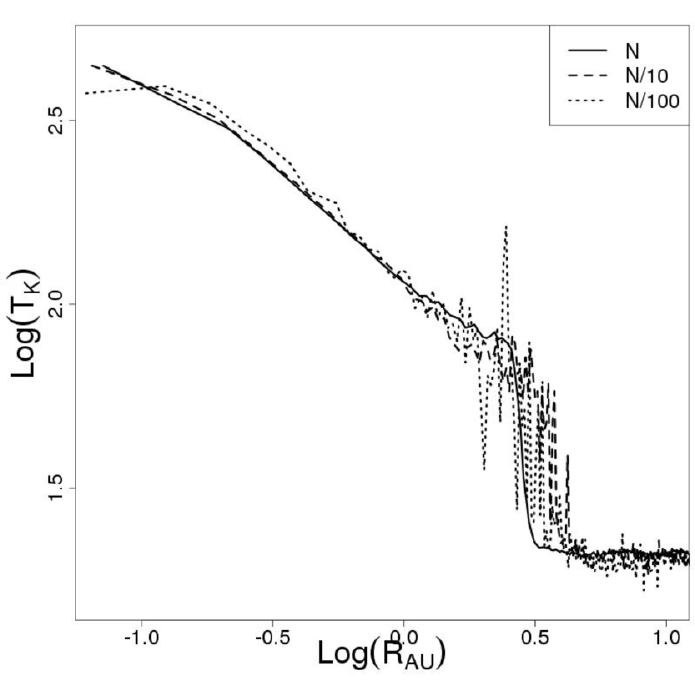}}
\caption{Temperature profiles for runs at different resoltions and at different times: from left to right $t= 1 T_{dyn}$,  $t= 1.5 T_{dyn}$,  $t= 3.0 T_{dyn}$. As in the previous figure, N is the number of particles in the high-resolution case.}
\label{fig:treso}
\end{figure}

\begin{figure}
\centering
\subfloat{\includegraphics[width = 0.5\textwidth]{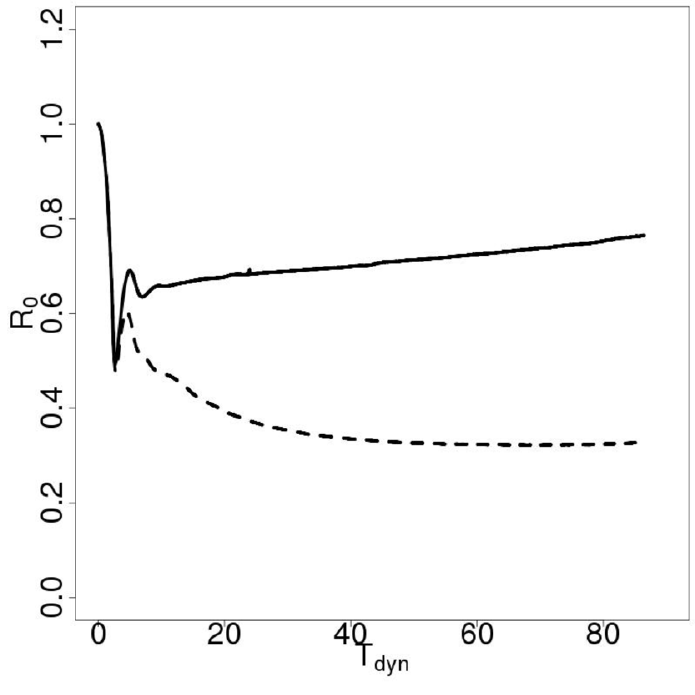}}
\subfloat{\includegraphics[width = 0.5\textwidth]{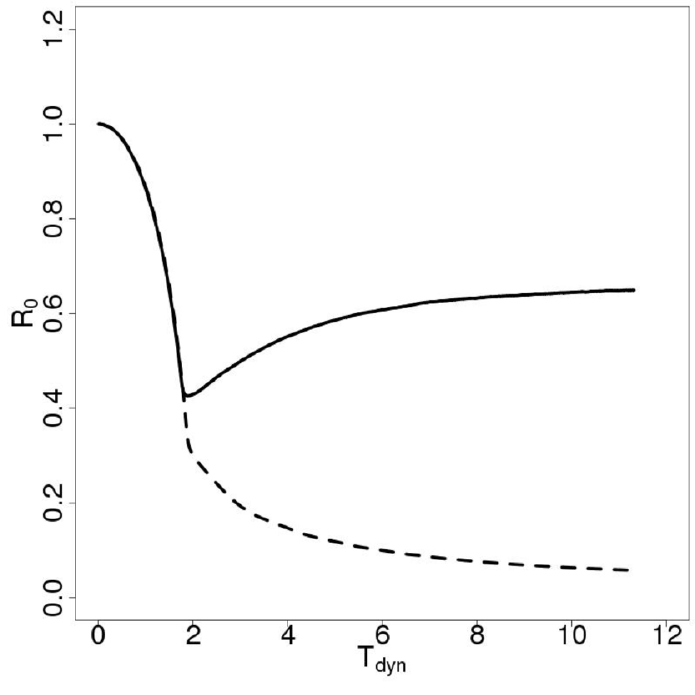}}
\caption{Evolution (in dynamical times) of the half-mass radius (in
  unit of its initial value) for IC$1$ (left) and IC$2$
  (right) in adiabatic (solid line) and cooling (dashed line) cases.}
\label{fig:hmr}
\end{figure}

\clearpage 

\begin{figure}
\centering
\includegraphics[width = 0.8\textwidth]{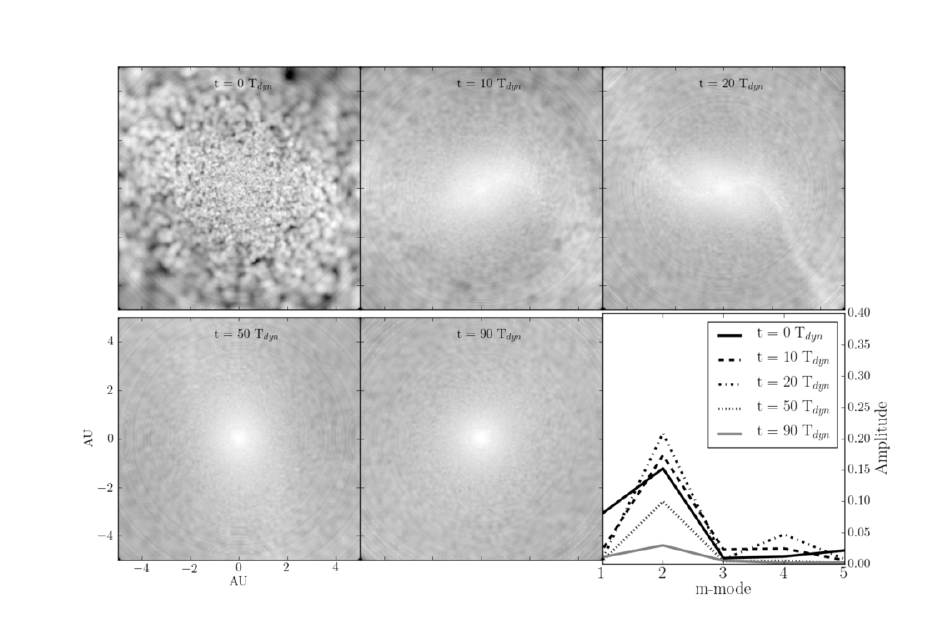}
\caption{Fourier analysis of the evolution of IC$1$ at different times: adiabatic case. The maps show the difference between the density and the mean density calculated from the surface density profile of the clump. Axes are in AU, see the bottom left corner. On the bottom right, corresponding amplitude vs Fourier modes}
\label{fig:m2a}
\end{figure}

\begin{figure}
\centering
\includegraphics[width = 0.8\textwidth]{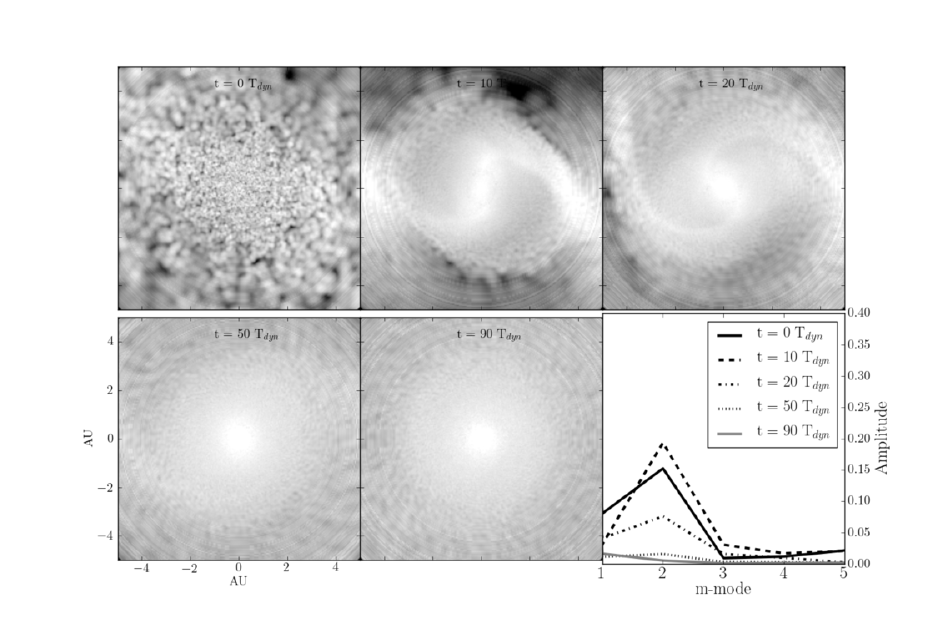}
\caption{Fourier analysis of the evolution of IC$1$ at different times: cooling case. The maps show the difference between the density and the mean density calculated from the surface density profile of the clump. Axes are in AU, see the bottom left corner. On the bottom right, corresponding amplitude vs Fourier modes}
\label{fig:m2c}
\end{figure}

\clearpage 

\begin{figure}
\centering
\subfloat{\includegraphics[width = 0.5\textwidth]{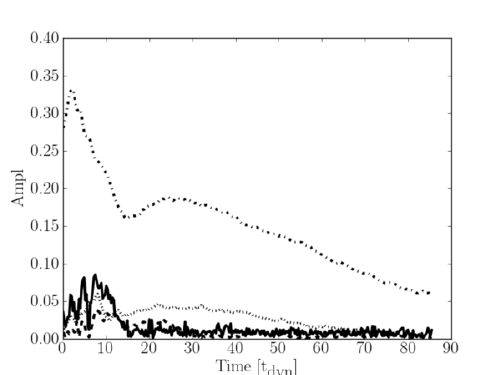}}
\subfloat{\includegraphics[width = 0.5\textwidth]{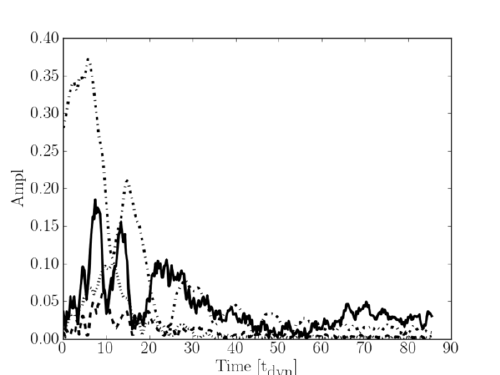}}
\caption{Evolution (in dynamical times) of the amplitude of the first four fourier modes in the adiabatic (left) and cooling (right) cases using IC$1$. The solid line corresponds to m=1, the dotted-dashed line to m=2, the dashed line to m=3 and the dotted line to m=4.}
\label{fig:mtime}
\end{figure}

\begin{figure}
\centering
\subfloat{\includegraphics[width = 0.5\textwidth]{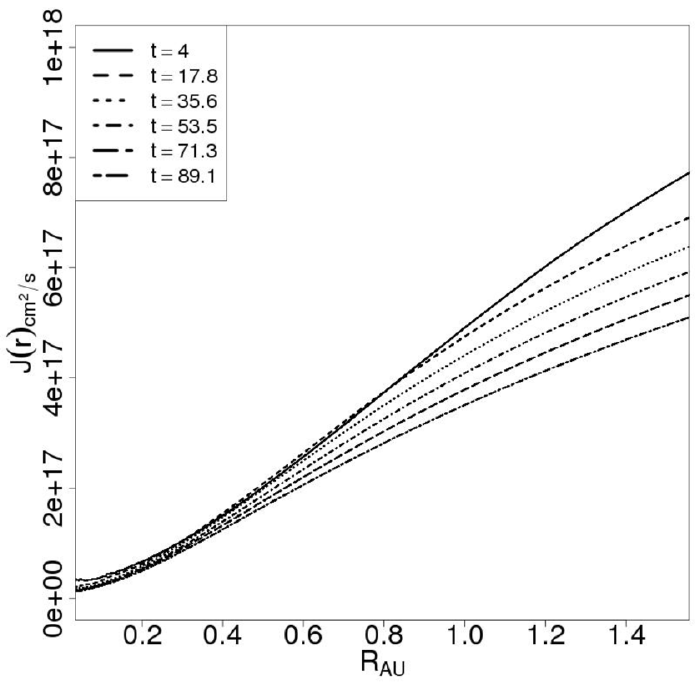}}
\subfloat{\includegraphics[width = 0.5\textwidth]{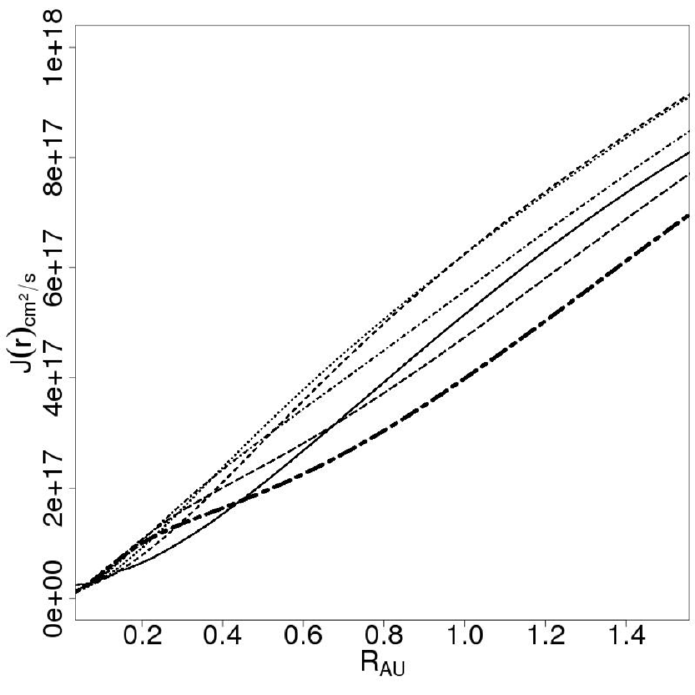}}
\caption{Evolution of the cumulative angular momentum profile in the
  adiabatic (left) and cooling (right) case for IC$1$. The transient phase (earlier than $4 t_{\mbox{dyn}}$) is not shown.}
\label{fig:sp}
\end{figure}

\begin{figure}
\centering
\subfloat{\includegraphics[width = 0.5\textwidth]{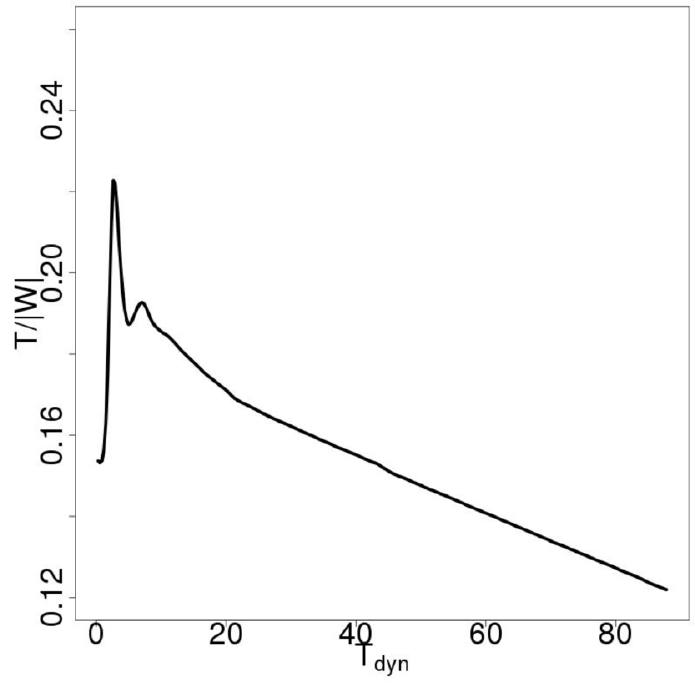}}
\subfloat{\includegraphics[width = 0.5\textwidth]{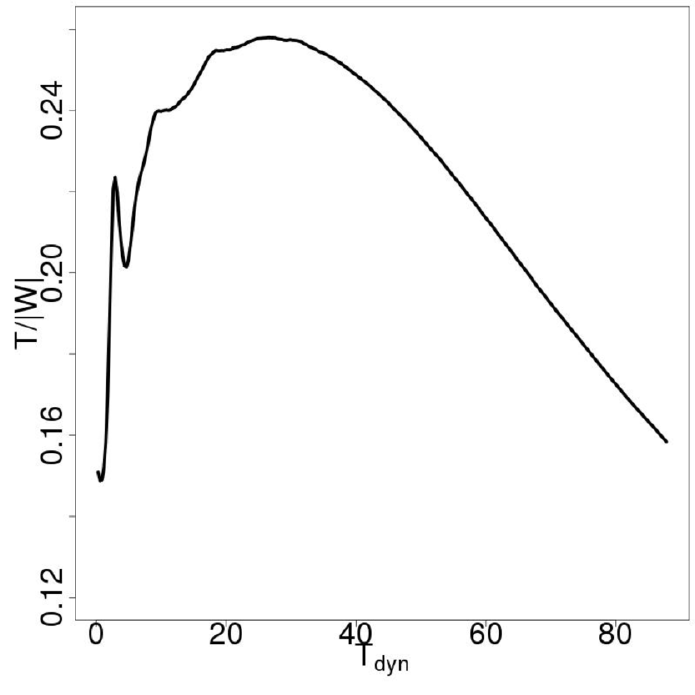}}
\caption{Evolution of the ratio between rotational and gravitational energy T/$\vert\vert W \vert\vert$ for IC$1$ in the adiabatic (on the left) and cooling (on the right) cases.}
\label{fig:tw}
\end{figure}

\begin{figure}
\centering
\includegraphics[width = 0.4\textwidth]{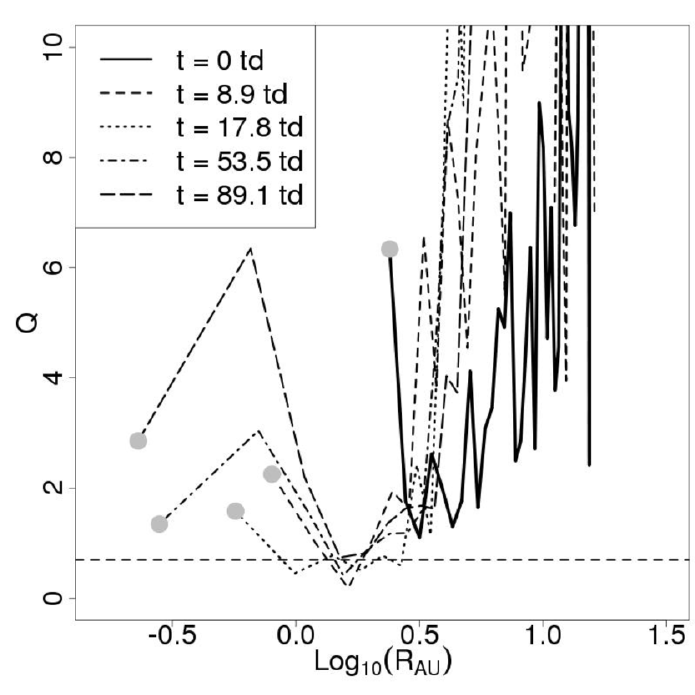}
\caption{Toomre parameter Q as a function of radius (in log scale) at five different times for the cooling evolution of IC$1$. The horizontal line is the threashold $0.7$ for a thick, $3$D disc (see, e.g., Mayer et a. $2004$).}
\label{fig:toom}
\end{figure}

\begin{figure}
\centering
\subfloat{\includegraphics[width = 0.35\textwidth]{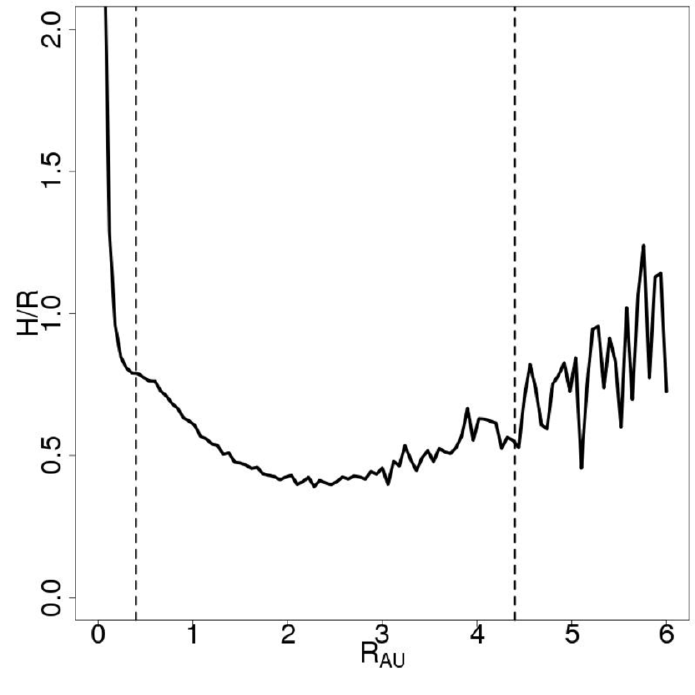}}
\subfloat{\includegraphics[width = 0.35\textwidth]{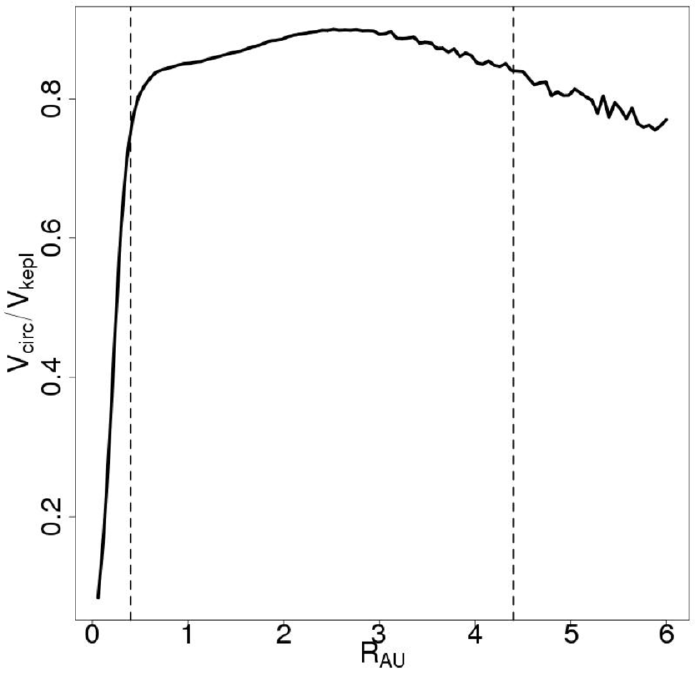}}
\subfloat{\includegraphics[width = 0.35\textwidth]{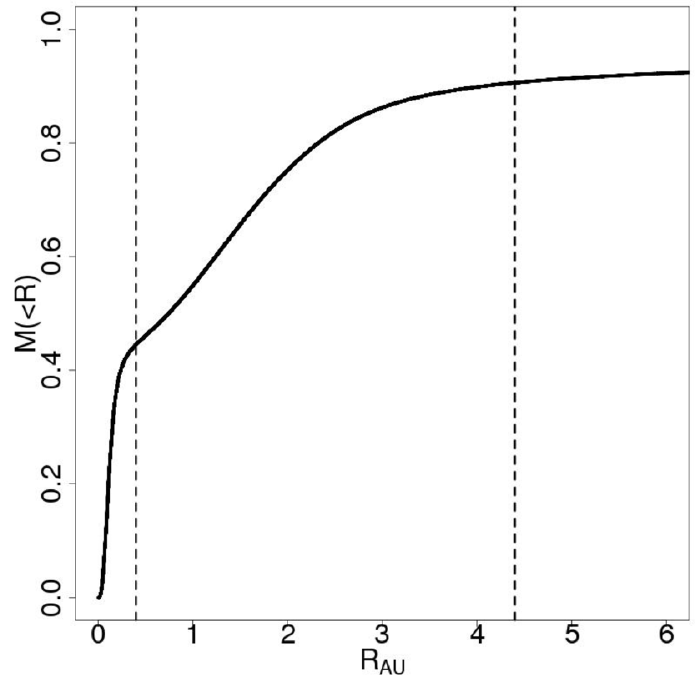}}
\caption{Circumplanetary disc properties at the last stage of IC$1$ simulation in cooling case. The left panel shows the ratio between the disc hight and the disc radius, the central panel the gas circular speed to the Keplerian speed, and the right panel the mass enclosed within a given radius. All profiles are shown versus clump radius. In each panel, the vertical lines represent the clump core and the disc boundaries.}
\label{fig:disc}
\end{figure}

\begin{figure}
\centering
\subfloat{\includegraphics[width = 0.4\textwidth]{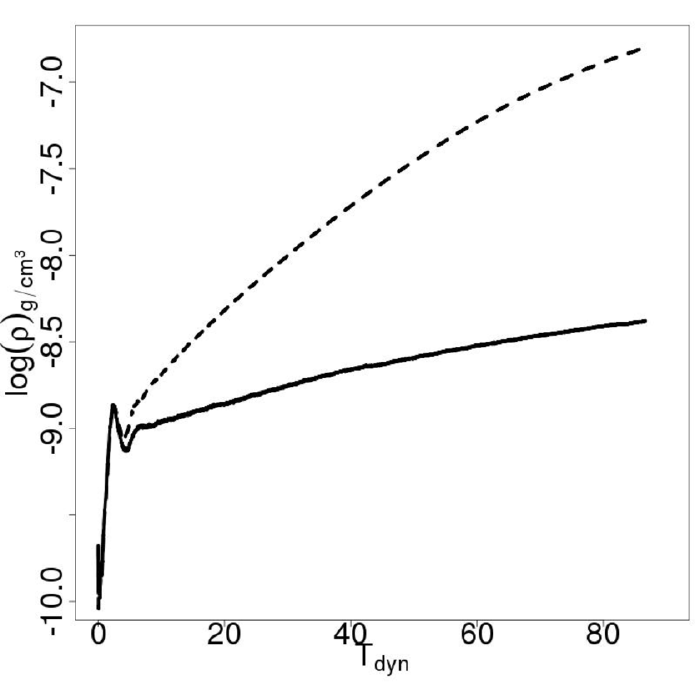}}
\subfloat{\includegraphics[width = 0.4\textwidth]{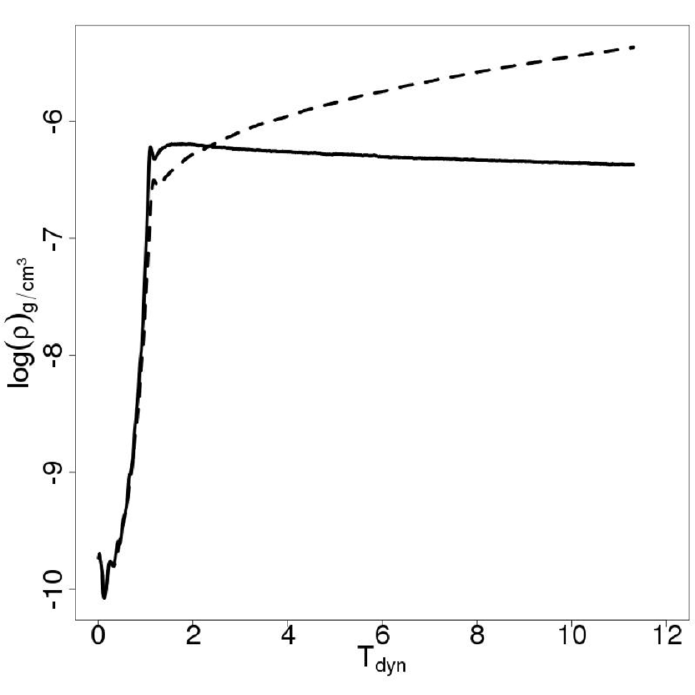}}
\caption{Evolution (in dynamical times) of the inner density (log
  scale of cgs units) for IC$1$ (left) and IC$2$
  (right) in adiabatic (solid line) and cooling (dashed line) cases.}
\label{fig:dens}
\end{figure}

\begin{figure}
\centering
\subfloat{\includegraphics[width = 0.4\textwidth]{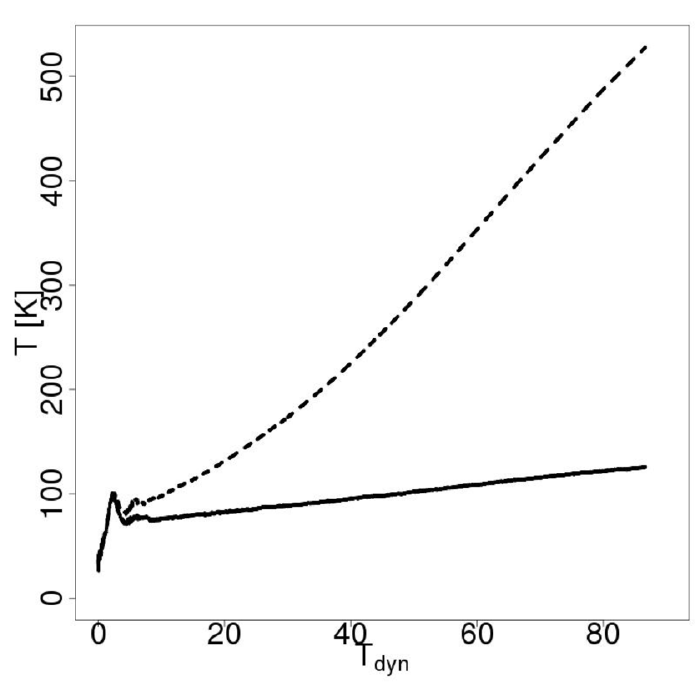}}
\subfloat{\includegraphics[width = 0.4\textwidth]{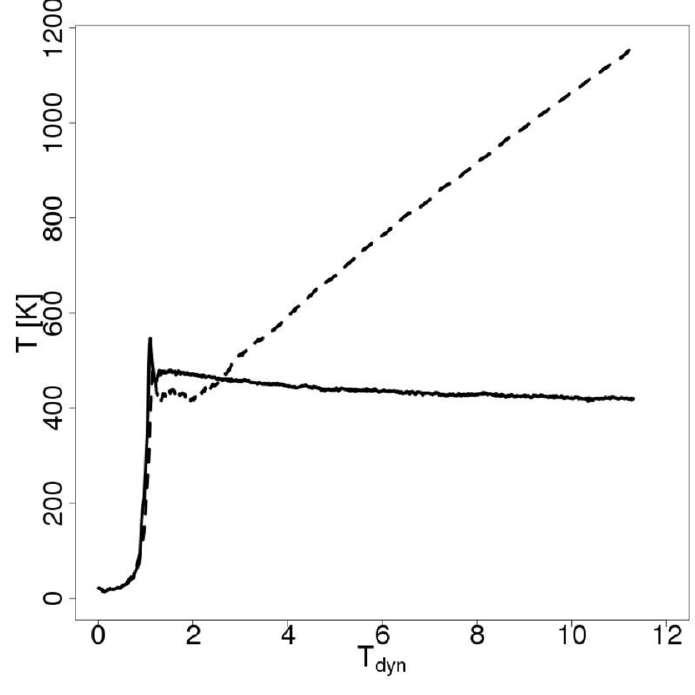}}
\caption{Evolution (in dynamical times) of the inner temperature for
 IC$1$ (left) and IC$2$ (right) in adiabatic (solid line) and cooling (dashed line) cases.}
\label{fig:temp}
\end{figure}

\begin{figure}
\centering
\subfloat{\includegraphics[width = 0.4\textwidth]{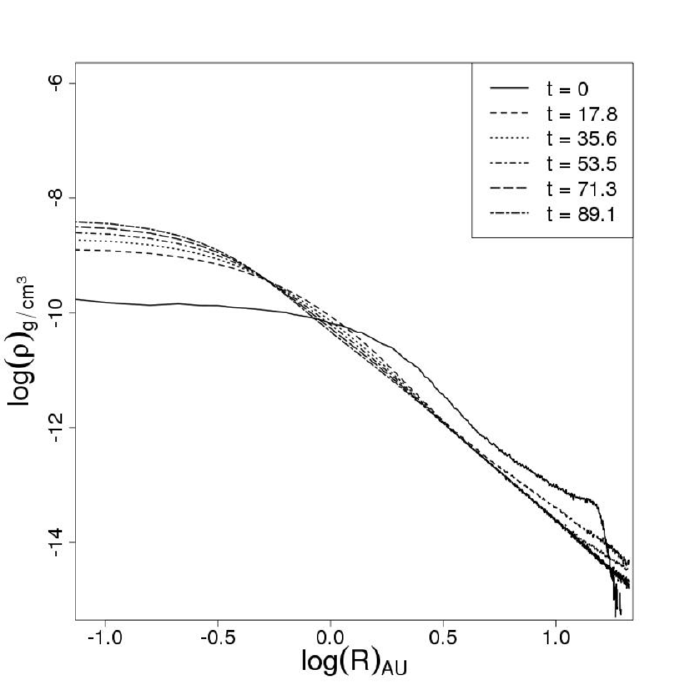}}
\subfloat{\includegraphics[width = 0.4\textwidth]{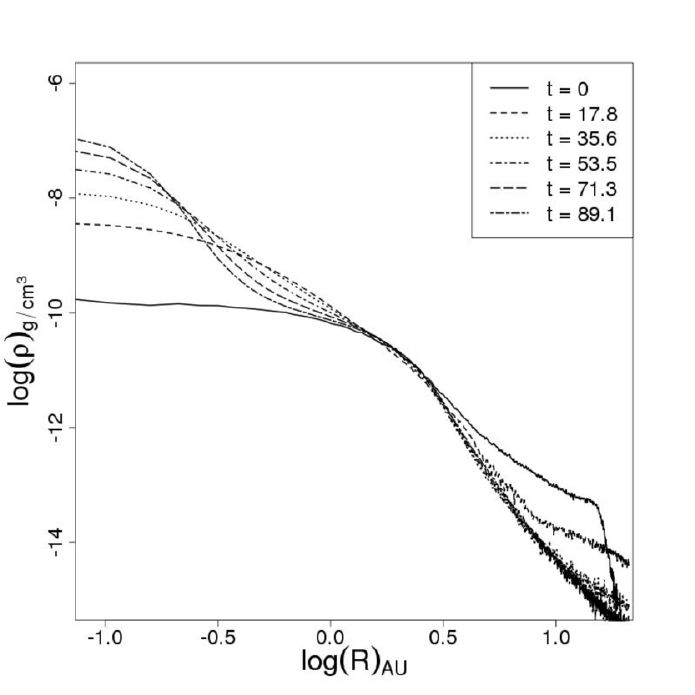}}
\caption{Density profile evolution for IC$1$ in the adiabatic (on the left) and cooling
  (on the right) cases.}
\label{fig:densE}
\end{figure}

\begin{figure}
\centering
\subfloat{\includegraphics[width = 0.4\textwidth]{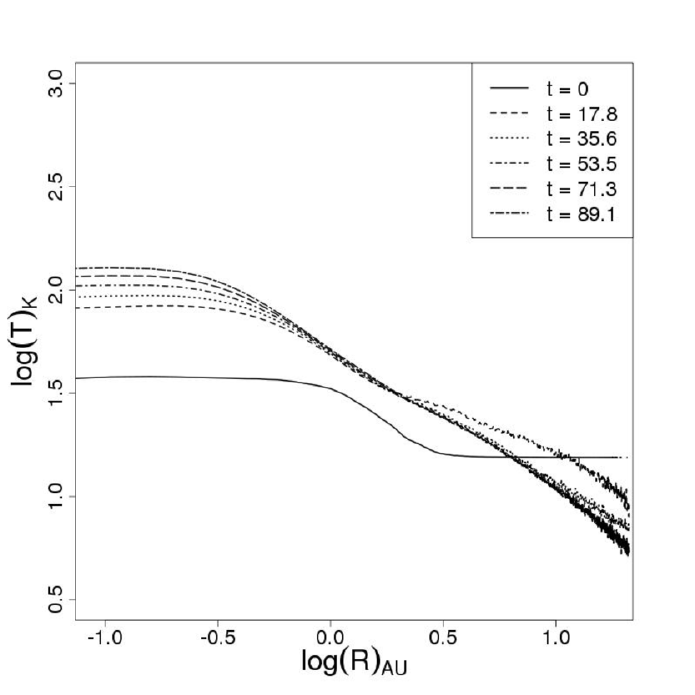}}
\subfloat{\includegraphics[width = 0.4\textwidth]{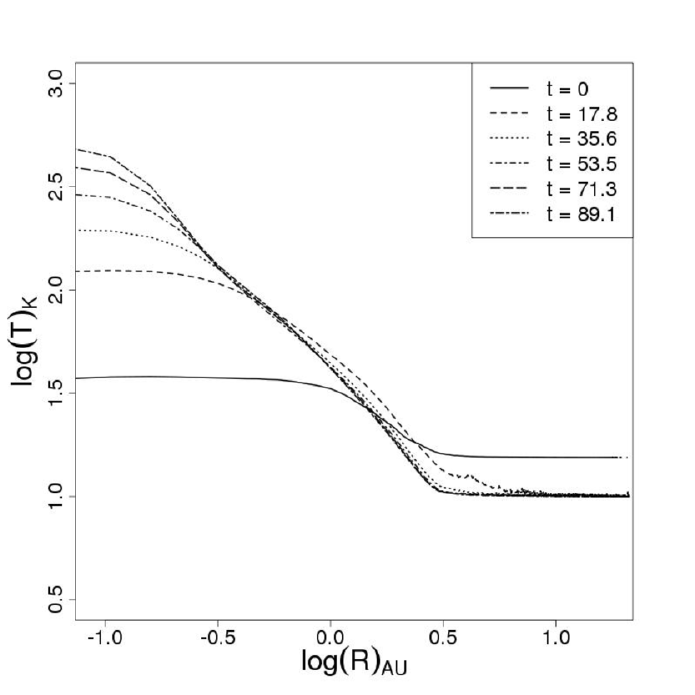}}
\caption{Temperature profile evolution for IC$1$ in the adiabatic (left) and cooling (right) cases.}
\label{fig:tempE}
\end{figure}

\begin{figure}
\centering
\subfloat{\includegraphics[width = 0.4\textwidth]{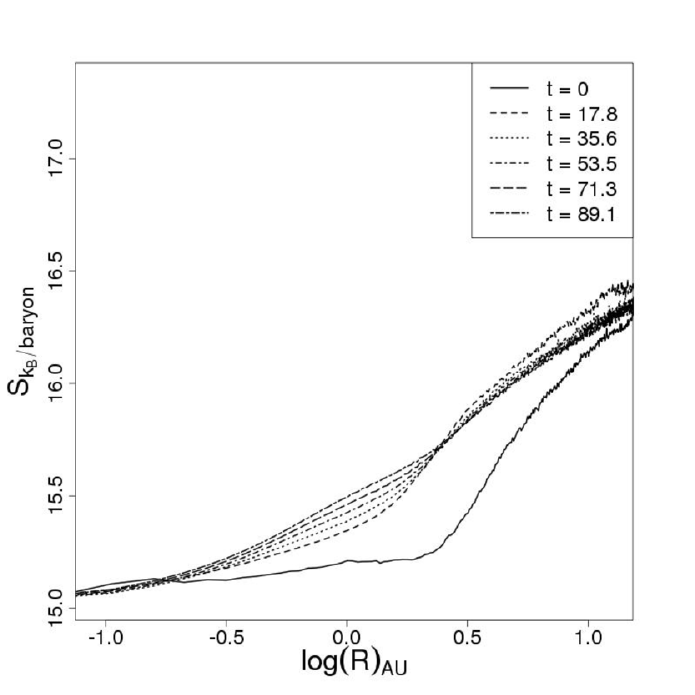}}
\subfloat{\includegraphics[width = 0.4\textwidth]{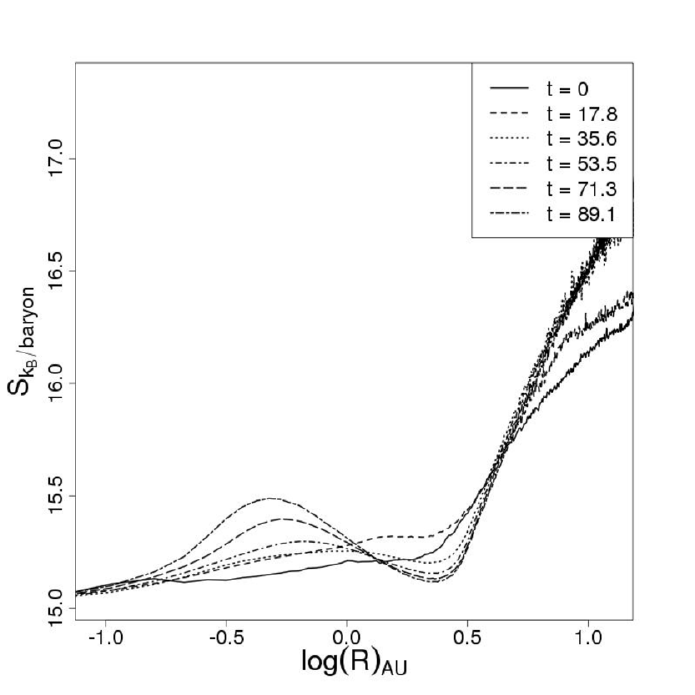}}
\caption{Specific entropy profile evolution for IC$1$ in the adiabatic (on the left) and cooling
  (on the right) cases.}
\label{fig:entrE}
\end{figure}

\newpage
\clearpage

\appendix
\def\mbox#1{{\rm#1}}
\def\summ{{\mathcal S}}
\def\Hatom{{\rm H}}
\def\Hmol{{\rm H_2}}

\section{Equation of State}

In this Appendix we derive the equation of state for a mixture of
atomic and diatomic hydrogen.  We define the dissociation fraction $a$ as
\begin{equation}
a = \frac{N_\Hatom}{N_p},
\end{equation}
with $N_\Hatom$ number of atoms and $N_p$ total number of protons.

The atomic and molecular parts are assumed to be in chemical
equilibrium.  Within the molecular part, however, the two nuclear-spin
configurations (ortho and para hydrogen) are not in equilibrium, but
taken to be frozen at $\rm ortho:para = 3:1$, as there is no efficient
mechanism to convert them into each other in dusty clumps \citep{boleyopr}.

\subsection*{Atomic component}

We now derive the thermodynamic functions for the atomic component.
The partition function will be made up of only the translational
component $q_t$ per particle
\begin{equation}
q_t = \frac{V}{\Lambda_{\mbox{at}}^3},
\end{equation}
where $V$ is the volume and $\Lambda_{\mbox{at}} = h / \sqrt{(2 \pi
  m_\Hatom k_B T)}$, with $h$ being Planck's constant, $m_\Hatom$ the
atomic mass of hydrogen, $k_B$ Boltzmann's constant and $T$ the
temperature. From that we can calculate $\log Q_\Hatom$:
\begin{equation}
\log Q_\Hatom = \log (q_t^{N_\Hatom})
\simeq N_\Hatom \log \frac{Ve}{N_\Hatom \Lambda_{\mbox{at}}^3}
= N_\Hatom \log \frac{e}{n_\Hatom \Lambda_{\mbox{at}}^3},
\end{equation} 
where $n_\Hatom$ is the number density. Now we can apply the
definitions for the Helmholtz energy $A_\Hatom$, internal energy
$E_\Hatom$, pressure $P_\Hatom$ and entropy $S_\Hatom$ to derive
\begin{eqnarray}
A_\Hatom & = &  -k_B T \log{ Q_\Hatom} =  -k_B T N_\Hatom
\log{\frac{e}{n \Lambda_{\mbox{at}}^3}} \\
E_\Hatom & = & k_B T^2
    \frac{\partial \log{Q_\Hatom}}{\partial T}\bigg|_{N_\Hatom, V}
  = \frac{3}{2} k_B N_\Hatom T \\
P_\Hatom & = & k_B T 
    \frac{\partial \log{Q_\Hatom}}{\partial V}\bigg|_{N_\Hatom, T}
  = \frac{N_\Hatom T k_B}{V} = \frac{k_B \rho T}{m_\Hatom} \\
S_\Hatom & = & \frac{E_\Hatom - A_\Hatom}{T}
  = N_\Hatom k_B \left( \frac{3}{2} +
    \log \frac{e}{n_\Hatom \Lambda_{\mbox{at}}^3} \right).
\end{eqnarray}

\subsection*{Diatomic component}

In the diatomic case, the partition function has a rotational
component $q_r$ and a vibrational component $q_v$.
\begin{eqnarray}
q_t & = &  \frac{V}{\Lambda_{\mbox{mol}}^3} \\
q_r & = & q_p^{1/4} \times q_o^{3/4} = \summ_E^{1/4} \summ_O^{3/4}
          \left( \exp{(2 \Theta_r/T)} \right)^{3/4}  \\
q_v & = & 1 / \left( 1- \exp{(-\Theta_v/T)} \right),
\end{eqnarray}
where $\Lambda_{\mbox{mol}} = h / \sqrt{(4 \pi m_\Hatom k_B T)}$,
$\Theta_r = 85 K$ is the critical rotational temperature and $\Theta_v
= 5987 K $ critical vibrational temperature, and 
\begin{eqnarray}
\summ_E &=& \sum_{j\;\mbox{even}} (2j+1) \exp{(-j(j+1)\Theta_r/T)}  \\
\summ_O &=& \sum_{j\;\mbox{odd}}  3(2j+1) \exp{(-j(j+1)\Theta_r/T)}  \,.
\end{eqnarray}
From this we can derive $\log{Q_\Hmol}$:
\begin{equation}
\frac{\log{Q_\Hmol}}{N_\Hmol} 
\simeq \frac{1}{4} \log \left( \summ_E \summ_O^3 \right)
+ \log{\frac{e}{\Lambda_{\mbox{mol}}^3 n_\Hmol}}
- \log{(1 - \exp{(- \Theta_v/T)})}  + \frac{3}{2}\frac{\Theta_r}{T}
\end{equation}

We can now derive the thermodynamical functions for diatomic hydrogen:
\begin{eqnarray*}
A_\Hmol & = & =  -k_B T N_\Hmol \left[   \frac{1}{4} \log{ \left( \summ_E \left( \summ_O \right)^3 \right) } + \frac{3}{2} \frac{\Theta_r}{T}  +  \log{\frac{e}{\Lambda_{\mbox{mol}}^3 n_\Hmol}} -\log{(1 - \exp{(- \Theta_v/T)})}    \right] \\
E_\Hmol & = & k_B N_\Hmol T \left[ \frac{3}{2} \left( 1 - \frac{\Theta_r}{T} \right)  +\frac{\Theta_v}{T} \frac{1}{\exp{\Theta_v/T}-1} + \frac{\Theta_r}{4T} \frac{ \overline{\summ_E}}{ \summ_E} + \frac{3 \Theta_r}{4T} \frac{ \overline{\summ_O}}{ \summ_O} \right] \\
P_\Hmol & = & k_B T \frac{\partial \log{Q_\Hmol}}{\partial V}|_{N_\Hmol, T} = \frac{N_\Hmol T k_B}{V} = k_B T n_\Hmol \\
S_\Hmol & = & N_\Hmol k_B  \left[ \frac{3}{2}  + \log{ \frac{e \summ_E^{1/4} \summ_O^{3/4} }{(1-e^{-\Theta_v/T})n_\Hmol \Lambda_{\mbox{mol}}^3}}+   \frac{\Theta_r}{4T} \left( \frac{\overline{\summ_E}}{ \summ_E} +3 \frac{\overline{\summ_O}}{ \summ_O }\right) + \frac{\Theta_v}{T}\frac{1}{e^{\Theta_v/T} - 1} \right],
\end{eqnarray*}
where
\begin{eqnarray*}
  \overline{\summ_E} & = & \sum_{j\;\mbox{even}} j(j+1)(2j+1) \exp{(-j(j+1)\Theta_r/T)} \\ 
  \overline{\summ_O} & = & 3 \sum_{j\;\mbox{odd}} j(j+1)(2j+1) \exp{(-j(j+1)\Theta_r/T)}.
\end{eqnarray*}

\subsection*{Dissociation fraction from chemical equilibrium}
We consider our system in chemical equilibrium, so that
\begin{equation}
2H \leftrightharpoons H_2.
\end{equation}
The value of the dissociation parameter $a$ is determined by the
equilibrium condition
\begin{equation}
\frac{[H]^2}{[H2]}=\frac{Q^2_\Hatom(\mbox{int})}{\Lambda_{\mbox{at}}^6}\frac{\Lambda_{\mbox{mol}}^3}{Q_\Hmol(\mbox{int})}
\end{equation}
The left side of the equation corresponds to the ratio between the
concentration of atomic / diatomic hydrogen:
\begin{equation}
\frac{[H]^2}{[H2]}= \frac{N^2_\Hatom}{V^2} \frac{V}{N_\Hmol} = \frac{N^2_p a^2}{V^2} \frac{2V}{(1-a)N_p} = \frac{2 a^2 N_p}{V(1-a)} = \frac{2a^2 n}{(1-a)},
\end{equation}
where $n$ is the total number density. For the right side of the
equation, we recall the functions used above, taking into account that
$Q_i(\mbox{int})$ is the total internal partition function, which does
not consider the translational term and instead has an extra term
\begin{equation}
\exp{( -E_g/(k_B T))}
\end{equation}
with $E_g$ ground state energy. Putting everything together, we find
that the equilibrium condition is
\begin{equation}
\frac{2a^2}{(1-a)} = \frac{F(T)}{n}
\end{equation}
with
\begin{equation}
F(T) = \left( \frac{\sqrt{(\pi m_\Hatom k_B T)}}{h}  \right)^3 e^{(-T_{\mbox{dis}}/(T))} \frac{1-e^{(-\Theta_V/T)}}{ \summ_E^{1/4} \summ_O^{3/4}
\exp{(3 \Theta_r / (2T))}},
\end{equation}
where $E_{\mbox{dis}}$ is the dissociation energy. The dissociation
fraction as a function of temperature and number density is
\begin{equation}
a(T,n) = \frac{1}{4} \left( - \frac{F(T)}{n} + \sqrt{\frac{F(T)^2}{n^2} + 8 \frac{F(T)}{n}}  \right).
\end{equation}

\subsection*{The equation of state}
As we know that the pressure is additive, we can find the total
pressure $P$ as
\begin{equation}
P = P_\Hatom + P_\Hmol = \frac{N_\Hatom T k_B}{V}  + \frac{N_\Hmol T k_B}{V} = \frac{k_B T}{V} a N_p \frac{(1-a) N_p}{2} = \frac{k_B T n(1+a)}{2}.
\end{equation}

\newpage 

\def\mnras{MNRAS}
\def\icarus{Icarus}
\def\apj{ApJ}
\def\aap{A \& A}
\def\apjl{ApJL}
\def\pasp{PASP}
\def\na{Nature}

\bibliographystyle{mn2e}
\bibliography{bibliomia}

\begin{thebibliography}{48}
\expandafter\ifx\csname natexlab\endcsname\relax\def\natexlab#1{#1}\fi

\bibitem[{{Armitage}(2010)}]{armitage}
{Armitage} P.~J., 2010, {Astrophysics of Planet Formation}

\bibitem[{{Ayliffe} \& {Bate}(2009)}]{2009MNRAS.397..657A}
{Ayliffe} B.~A., {Bate} M.~R., 2009, \mnras, 397, 657

\bibitem[{{Baruteau} {et~al}\mbox{.}(2011){Baruteau}, {Meru}, \&
  {Paardekooper}}]{2011MNRAS.416.1971B}
{Baruteau} C., {Meru} F., {Paardekooper} S.-J., 2011, \mnras, 416, 1971

\bibitem[{{Boley}(2009)}]{boley2009}
{Boley} A.~C., 2009, \apjl, 695, L53

\bibitem[{{Boley} \& {Durisen}(2010)}]{2010ApJ...724..618B}
{Boley} A.~C., {Durisen} R.~H., 2010, \apj, 724, 618

\bibitem[{{Boley} {et~al}\mbox{.}(2007){Boley}, {Hartquist}, {Durisen}, \&
  {Michael}}]{boleyopr}
{Boley} A.~C., {Hartquist} T.~W., {Durisen} R.~H., {Michael} S., 2007, \apjl,
  656, L89

\bibitem[{{Boley} {et~al}\mbox{.}(2010){Boley}, {Hayfield}, {Mayer}, \&
  {Durisen}}]{2010Icar..207..509B}
{Boley} A.~C., {Hayfield} T., {Mayer} L., {Durisen} R.~H., 2010, \icarus, 207,
  509

\bibitem[{{Bonavita} {et~al}\mbox{.}(2011){Bonavita}, {Jayawardhana}, {Janson},
  \& {Lafreni{\`e}re}}]{bonavita}
{Bonavita} M., {Jayawardhana} R., {Janson} M., {Lafreni{\`e}re} D., 2011, in
  IAU Symposium, Vol. 276, IAU Symposium, {A.Sozzetti, M.G.Lattanzi, \&
  A.P.Boss}, ed., pp. 113--116

\bibitem[{{Borucki} {et~al}\mbox{.}(2011){Borucki}, {Koch}, {Basri}, {Batalha},
  {Boss}, {Brown}, \& {Caldwell}}]{2011ApJ...728..117B}
{Borucki} W.~J., {Koch} D.~G., {Basri} G., {Batalha} N., {Boss} A., {Brown}
  T.~M., {Caldwell}, 2011, \apj, 728, 117

\bibitem[{{Boss}(1997)}]{1997Sci...276.1836B}
{Boss} A.~P., 1997, Science, 276, 1836

\bibitem[{{Desch}(2007)}]{2007ApJ...671..878D}
{Desch} S.~J., 2007, \apj, 671, 878

\bibitem[{{Durisen} {et~al}\mbox{.}(2007){Durisen}, {Boss}, {Mayer}, {Nelson},
  {Quinn}, \& {Rice}}]{2007prpl.conf..607D}
{Durisen} R.~H., {Boss} A.~P., {Mayer} L., {Nelson} A.~F., {Quinn} T., {Rice}
  W.~K.~M., 2007, Protostars and Planets V, 607

\bibitem[{{Durisen} {et~al}\mbox{.}(1986){Durisen}, {Gingold}, {Tohline}, \&
  {Boss}}]{1986ApJ...305..281D}
{Durisen} R.~H., {Gingold} R.~A., {Tohline} J.~E., {Boss} A.~P., 1986, \apj,
  305, 281

\bibitem[{{Gammie}(2001)}]{2001ApJ...553..174G}
{Gammie} C.~F., 2001, \apj, 553, 174

\bibitem[{{Hashimoto} {et~al}\mbox{.}(2011){Hashimoto}, {Tamura}, {Muto},
  {Kudo}, {Fukagawa}, {Fukue}, {Goto}, {Grady}, {Henning}, {Hodapp}, {Honda},
  {Inutsuka}, {Kokubo}, {Knapp}, \& {McElwain}}]{2011ApJ...729L..17H}
{Hashimoto} J. {et~al.}, 2011, \apjl, 729, L17

\bibitem[{{Helled} \& {Bodenheimer}(2011)}]{2011Icar..211..939H}
{Helled} R., {Bodenheimer} P., 2011, \icarus, 211, 939

\bibitem[{{Helled} \& {Schubert}(2008)}]{helled}
{Helled} R., {Schubert} G., 2008, \icarus, 198, 156

\bibitem[{{Helled} \& {Schubert}(2009)}]{2009ApJ...697.1256H}
{Helled} R., {Schubert} G., 2009, \apj, 697, 1256

\bibitem[{{Kuiper}(1951)}]{1951PNAS...37....1K}
{Kuiper} G.~P., 1951, Proceedings of the National Academy of Science, 37, 1

\bibitem[{{Lafreni{\`e}re} {et~al}\mbox{.}(2010){Lafreni{\`e}re},
  {Jayawardhana}, \& {van Kerkwijk}}]{2010ApJ...719..497L}
{Lafreni{\`e}re} D., {Jayawardhana} R., {van Kerkwijk} M.~H., 2010, \apj, 719,
  497

\bibitem[{{Lodato} \& {Clarke}(2011)}]{2011MNRAS.413.2735L}
{Lodato} G., {Clarke} C.~J., 2011, \mnras, 413, 2735

\bibitem[{{Machida} {et~al}\mbox{.}(2008){Machida}, {Kokubo}, {Inutsuka}, \&
  {Matsumoto}}]{machidaAM}
{Machida} M.~N., {Kokubo} E., {Inutsuka} S.-i., {Matsumoto} T., 2008, \apj,
  685, 1220

\bibitem[{{Martin} \& {Lubow}(2011)}]{2011MNRAS.413.1447M}
{Martin} R.~G., {Lubow} S.~H., 2011, \mnras, 413, 1447

\bibitem[{{Mayer} {et~al}\mbox{.}(2002){Mayer}, {Quinn}, {Wadsley}, \&
  {Stadel}}]{mayerquinn}
{Mayer} L., {Quinn} T., {Wadsley} J., {Stadel} J., 2002, Science, 298, 1756

\bibitem[{{Mayer} {et~al}\mbox{.}(2004){Mayer}, {Quinn}, {Wadsley}, \&
  {Stadel}}]{2004ApJ...609.1045M}
{Mayer} L., {Quinn} T., {Wadsley} J., {Stadel} J., 2004, \apj, 609, 1045

\bibitem[{{Meru} \& {Bate}(2011)}]{2011MNRAS.411L...1M}
{Meru} F., {Bate} M.~R., 2011, \mnras, 411, L1

\bibitem[{{Michael} {et~al}\mbox{.}(2011){Michael}, {Durisen}, \&
  {Boley}}]{2011ApJ...737L..42M}
{Michael} S., {Durisen} R.~H., {Boley} A.~C., 2011, \apjl, 737, L42

\bibitem[{{Mizuno}(1980)}]{mizuno}
{Mizuno} H., 1980, Progress of Theoretical Physics, 64, 544

\bibitem[{{Movshovitz} {et~al}\mbox{.}(2010){Movshovitz}, {Bodenheimer},
  {Podolak}, \& {Lissauer}}]{2010Icar..209..616M}
{Movshovitz} N., {Bodenheimer} P., {Podolak} M., {Lissauer} J.~J., 2010,
  \icarus, 209, 616

\bibitem[{{Muto} {et~al}\mbox{.}(2012){Muto}, {Grady}, {Hashimoto}, {Fukagawa},
  {Hornbeck}, {Sitko}, {Russell}, {Werren}, {Cur{\'e}}, \&
  {Currie}}]{2012ApJ...748L..22M}
{Muto} T. {et~al.}, 2012, \apjl, 748, L22

\bibitem[{{Nayakshin}(2010)}]{nayakshin}
{Nayakshin} S., 2010, \mnras, 408, L36

\bibitem[{{Nielsen} {et~al}\mbox{.}(2011){Nielsen}, {Liu}, {Wahhaj}, {Biller},
  {Chun}, {Close}, {Ftaclas}, {Hayward}, {Toomey}, \& {Gemini NICI
  Planet-Finding Campaign Team}}]{gemini}
{Nielsen} E.~L. {et~al.}, 2011, American Astronomical Society, ESS meeting, 2,
  702

\bibitem[{{Paardekooper}(2012)}]{2012MNRAS.421.3286P}
{Paardekooper} S.-J., 2012, \mnras, 421, 3286

\bibitem[{{Perna} {et~al}\mbox{.}(2010){Perna}, {Menou}, \&
  {Rauscher}}]{rosalba}
{Perna} R., {Menou} K., {Rauscher} E., 2010, \apj, 719, 1421

\bibitem[{{Pollack} {et~al}\mbox{.}(1996){Pollack}, {Hubickyj}, {Bodenheimer},
  {Lissauer}, {Podolak}, \& {Greenzweig}}]{pollack}
{Pollack} J.~B., {Hubickyj} O., {Bodenheimer} P., {Lissauer} J.~J., {Podolak}
  M., {Greenzweig} Y., 1996, \icarus, 124, 62

\bibitem[{{Quanz} {et~al}\mbox{.}(2012){Quanz}, {Lafreniere}, {Meyer},
  {Reggiani}, \& {Buenzli}}]{quanz}
{Quanz} S.~P., {Lafreniere} D., {Meyer} M.~R., {Reggiani} M.~M., {Buenzli} E.,
  2012, ArXiv e-prints

\bibitem[{{Rafikov}(2009)}]{rafikov2009}
{Rafikov} R.~R., 2009, \apj, 704, 281

\bibitem[{{Rafikov}(2011)}]{2011ApJ...727...86R}
{Rafikov} R.~R., 2011, \apj, 727, 86

\bibitem[{{Rice} {et~al}\mbox{.}(2005){Rice}, {Lodato}, \&
  {Armitage}}]{2005MNRAS.364L..56R}
{Rice} W.~K.~M., {Lodato} G., {Armitage} P.~J., 2005, \mnras, 364, L56

\bibitem[{{Rogers} \& {Wadsley}(2012)}]{rogers_wadsley_2012_mnras}
{Rogers} P.~D., {Wadsley} J., 2012, \mnras, 2995

\bibitem[{{Shakura} \& {Sunyaev}(1973)}]{shakura_sunyaev_1973}
{Shakura} N.~I., {Sunyaev} R.~A., 1973, \aap, 24, 337

\bibitem[{{Spiegel} \& {Burrows}(2012)}]{burrows}
{Spiegel} D.~S., {Burrows} A., 2012, \apj, 745, 174

\bibitem[{{Stadel}(2001)}]{Stadel:2001}
{Stadel} J.~G., 2001, PhD thesis, University of Washington

\bibitem[{{Stamatellos} {et~al}\mbox{.}(2007){Stamatellos}, {Whitworth},
  {Bisbas}, \& {Goodwin}}]{2007A&A...475...37S}
{Stamatellos} D., {Whitworth} A.~P., {Bisbas} T., {Goodwin} S., 2007, \aap,
  475, 37

\bibitem[{{Sumi} {et~al}\mbox{.}(2011){Sumi}, {MOA}, \& {OGLE
  Collaboration}}]{2011ESS.....2.0103S}
{Sumi} T., {MOA}, {OGLE Collaboration}, 2011, American Astronomical Society,
  ESS meeting, 2, 103

\bibitem[{{Takata} \& {Stevenson}(1996)}]{takata}
{Takata} T., {Stevenson} D.~J., 1996, \icarus, 123, 404

\bibitem[{{Wadsley} {et~al}\mbox{.}(2004){Wadsley}, {Stadel}, \&
  {Quinn}}]{GasolineW}
{Wadsley} J.~W., {Stadel} J., {Quinn} T., 2004, \na, 9, 137

\bibitem[{{Wright} {et~al}\mbox{.}(2012){Wright}, {Marcy}, {Howard}, {Johnson},
  {Morton}, \& {Fischer}}]{2012arXiv1205.2273W}
{Wright} J.~T., {Marcy} G.~W., {Howard} A.~W., {Johnson} J.~A., {Morton} T.,
  {Fischer} D.~A., 2012, ArXiv e-prints

\end{thebibliography}

\end{document}